\documentclass{article}
\usepackage[textwidth=1.2\textwidth]{geometry}
\usepackage[utf8]{inputenc}
\usepackage[T1]{fontenc}
\usepackage[final]{fixme}
\usepackage{natbib}
\usepackage{array}
\usepackage{afterpage}
\usepackage{comment}
\usepackage{randtext}
\usepackage{stmaryrd} 
\usepackage{upgreek}
\usepackage{enumitem}
\usepackage{amsmath}
\usepackage{amssymb}
\usepackage{eqntabular,renumber,calculus}
\usepackage{graphicx}
\graphicspath{{../invariance/}{../cat-semantics/}}
\usepackage{relsize}
\usepackage{phaistos}
\usepackage{multicol}
\usepackage{placeins}
\usepackage[british]{babel}
\usepackage{tikz}
\input{macros.sty}
\usepackage{LISA}
\usepackage{interval}

\usepackage{enumitem}
\usepackage{xspace}
\makeatletter

\newcommand{\nbsp}{{\mbox{$~$}}}
\newif\ifitalicenv\italicenvtrue

\newtheorem{theorem}{Theorem}[section]

\newtheorem{exam}[theorem]{Example}

\newtheorem{defi}[theorem]{Definition}

\def\@begintheorem#1#2{%
    \trivlist
    \item[%
        \hskip 12\p@
        \hskip \labelsep
        {\ifitalicenv\sc\else\itshape\fi #1\hskip 5\p@\relax{\rm #2}.\enspace}]%
        \ifitalicenv\itshape\else\upshape\fi\hskip-\labelsep%
}
\def\@opargbegintheorem#1#2#3{%
    \trivlist
    \item[\hskip 12pt
          \hskip \labelsep
          {\ifitalicenv{\sc{#1}}\else{\itshape#1}\fi%
	   \savebox\@tempboxa{\ifitalicenv{\scshape#3}\else{\itshape#3}\fi}%
	   \ifdim\wd\@tempboxa>\z@%
           \ {\rm #2}\unskip\hskip5pt\relax$($\box\@tempboxa$)$%
	   \fi.\unskip\hskip5pt}]
\ifitalicenv\itshape\else\upshape\fi\hskip-\labelsep}

\newif\if@qeded
\global\@qededfalse
\def\proof{%
    \global\@qededfalse
    \@ifnextchar[{\@xproof}{\@proof}}

\def\endproof{%
    \if@qeded\else\qed\fi
    \endtrivlist
}
\def\@proof{%
    \trivlist
    \item[%
        \hskip 12\p@
        \hskip \labelsep
        {\sc Proof.\enspace}]\hskip-\labelsep%
    \ignorespaces
}
\def\@xproof[#1]{%
    \trivlist
    \item[\hskip 12\p@\hskip \labelsep{\sc Proof #1.}]%
    \ignorespaces
}
\def\qed{\unskip\kern 10pt{\unitlength1pt\linethickness{.4pt}\framebox(5,5){}}
    \global\@qededtrue
    }%
\def\newdef#1#2{%
    \expandafter\@ifdefinable\csname #1\endcsname
        {\@definecounter{#1}%
         \expandafter\xdef\csname the#1\endcsname{\@thmcounter{#1}}%
         \global\@namedef{#1}{\@defthm{#1}{#2}}%
         \global\@namedef{end#1}{\@endtheorem}%
    }%
}
\def\@defthm#1#2{%
    \refstepcounter{#1}%
    \@ifnextchar[{\@ydefthm{#1}{#2}}{\@xdefthm{#1}{#2}}%
}
\def\@xdefthm#1#2{%
    \@begindef{#2}{\csname the#1\endcsname}%
    \ignorespaces
}
\def\@ydefthm#1#2[#3]{%
    \trivlist
    \item[%
        \hskip 10\p@
        \hskip \labelsep
        {\it #2%
         \savebox\@tempboxa{#3}%
         \ifdim \wd\@tempboxa>\z@
            \ \box\@tempboxa
         \fi.%
        }]%
    \ignorespaces
}
\def\@begindef#1#2{%
    \trivlist
    \item[%
        \hskip 10\p@
        \hskip \labelsep
        {\it #1\ \rm #2.}%
    ]%
}

\makeatother

\newcommand{\cattuple}{\quintuple}

\usepackage{hyperref}

\title{\textbf{\LARGE Syntax and semantics of the weak consistency model specification language \cat}}
\author{\begin{tabular}{c}
\Large\bfseries Jade Alglave \\[1ex]
Microsoft Research Cambridge \\ 
University College London\\
\texttt{\randomize{jaalglav@microsoft.com}}, \texttt{\randomize{j.alglave@ucl.ac.uk}}
\\[1em]
\Large\bfseries Patrick Cousot\\[1ex]
New York University\\ 
emer. \'Ecole Normale Sup\'erieure, PSL Research University\\
\texttt{\randomize{pcousot@cims.nyu.edu}}, \texttt{\randomize{cousot@ens.fr}}
\\[1em]
\Large\bfseries Luc Maranget\\[1ex]
INRIA\\
\texttt{\randomize{Luc.Maranget@inria.fr}}
\\[1ex]
\end{tabular}}

\begin{document}

\maketitle

\newcommand{\refsection}[1]{Sect.\nbsp\ref{#1}}
\newcommand{\reffigure}[1]{Fig.\nbsp\ref{#1}}

\begin{abstract}
We provide the syntax and semantics of the \cat language,  a domain specific language to describe consistency properties of parallel/distributed programs. The language is implemented in the \texttt{herd7} tool \cite{herd7}. 
\end{abstract}

\section{Introduction}

The \texttt{cat} language \cite{AlglaveCousotMaranget-cat-HSA-2015}
is a domain specific language to describe consistency properties succinctly by constraining an abstraction of parallel program executions into a candidate execution and possibly extending this candidate execution with additional constraints on the execution environment. The \emph{analytic semantics} of a program is defined by its \emph{anarchic semantics} that is a set of executions describing computations and a \emph{\cat specification} \catfile{}  describing a weak memory model. An example of anarchic semantics semantics for \LISA is given in \cite{LISA-arxiv16}. An anarchic semantics is a truly parallel semantics, with no global time, describing all possible computations with all possible communications. 
The \cat language operates on abstractions of the anarchic executions called \emph{candidate executions}. The \cat specification \catfile{} checks a candidate execution for the consistency specification (including, maybe, by defining constraints on the program execution environments, such as the the final writes or the coherence order).

The abstraction of an anarchic execution into a candidate execution is overview in Section \ref{sec:Abstraction-candidate-executions} while the \cat language is introduced is
Section \ref{sec:cat-language}. Its formal semantics is defined in Section \ref{sec:lisa-cat}. Examples can be found in Alglave [2015].

\section{Abstraction to candidate executions}\label{sec:Abstraction-candidate-executions}
The anarchic semantics is a set of executions. Each execution is abstracted to a 
candidate execution
$\cattuple{\setofevents}{\catpo}{\catrf}{\catIW}{\scoperelationitalique}$ 
providing
\begin{itemize}[align=left,leftmargin=*,topsep=0pt,itemsep=0pt,parsep=1pt]
\item \emph{events} $\setofevents$, giving a semantics to instructions; for example in \LISA{} \cite{LISA-arxiv16}, a write instruction  \texttt{w[] x v} yields a write event of variable \texttt{x} with value {\tt
v}. Events can be (for brevity this is not an exhaustive list):
\begin{itemize}[align=left,leftmargin=*,topsep=-5pt,itemsep=-2pt]
\item \emph{writes}, gathered in the set \catW, including the
the set $\catIW$ of \emph{initial writes}  
   coming from the
prelude of the program; 
  \item \emph{reads}, gathered in the set \catR;
  \item \emph{branch} events, gathered in the set \catB;
  \item \emph{fences}, gathered in the set \catF. 
  \end{itemize}
\item the program order \catpo, relating accesses written in program order in
the original \LISA program;
\item the read-from $\catrf$ describing a communication between a write and a read event;
\item the scope relation $\scoperelationitalique$ relating events that come from threads which reside within the same scope;
\end{itemize}
A \cat specification \catfile{} may add other components to the candidate execution (\eg to specify constraints on the execution environment) and then checks that this extended candidate execution satisfies the consistency specification, that is, essentially, that the communication relation $\catrf$ satisfies the consistency specification (under hypotheses on the execution environment).

\section{The \cat\ language}\label{sec:cat-language}

A weak consistency specification written in the \cat\ language defines
constraints to be satisfied by the communication relation $\catrf$ of any
candidate execution. A typical \cat\ specification defines new objects
depending on the sets and relations of the candidate execution (\eg{} the
program order {\tt po} or the initial writes {\tt IW}) and then imposes
constraints on these objects that ultimately restrict the allowed
communications $\catrf$ .

\subsection{Objects and expressions}

\subsubsection{Types.\quad}
The objects defined in a \cat\ specification may be of the following types (see Appendix~\ref{sec:types-semantic-domains} and Figure \ref{fig:sem-doms} for the formal details): 
$\cattypenocolon{evt}{}$ (event),
$ \cattypenocolon{tag}{}$ (tag),
$ \cattypenocolon{rel}{}$ (relation between events),
$ \cattypenocolon{set}{}$ (set),
$ \cattypenocolon{tuple}{}$ (tuple),
$ \cattypenocolon{enum}{}$ (enumeration of tags),
$ \cattypenocolon{fun}{}$ (unary function type),
$ \cattypenocolon{proc}{}$ (unary procedure type).

\subsubsection{Definitions in binding statements} (see
Appendix~\ref{sec:Binding-definitions-sem} and Figure \ref{fig:bind} for their
formal semantics) can bind an expression to a name, which can be used in place
of that expression. For example
\begin{quote}
\begin{verbatim}
let rfe = rf & ext
\end{verbatim}
\end{quote}
defines the relation {\tt rfe} as the restriction of the communications {\tt
rf} to events coming from different processes. Formally, {\tt rfe} is built as
the intersection (denoted by {\tt \&} in \cat{}) of the read-from relation {\tt
rf} and the predefined relation {\tt ext} which links events coming from different
processes (Figure \ref{fig:predef-rlns-sem}). 

A set, relation, function or procedure can be given a name by binding (see
Figure \ref{fig:definitions}). Bindings (see Appendix~\ref{sec:bindings-sem}
and Figure \ref{fig:bind}) can be (mutually) recursive (using \texttt{let rec
\ldots\ and \ldots}). 

\subsubsection{Functions}\label{sec:Functions-intro} (see
Appendix~\ref{sec:functions-sem} and Figure~\ref{fig:fun-sem} for their formal
semantics) define an object as a function of a unique formal parameter (which
may be an empty tuple \texttt{()} in absence of parameter or a non-empty tuple
for multiple parameters).  For example
\begin{quote}
\begin{verbatim}
let extof r = r & ext
let rfe = extof rf
\end{verbatim}
\end{quote}
defines a function {\tt extof} of a parameter {\tt r} which intersects the
relation {\tt r} with the relation {\tt ext} between events belonging to
different processes. We then define the relation {\tt rfe} as the function {\tt
extof} applied to the read-from relation {\tt rf}.

We note that our definition of {\tt extof} above  is an abbreviation for the
binding of an anonymous function
\begin{quote}
\begin{verbatim}
let extof = fun r -> r & ext
\end{verbatim}
\end{quote}
Functions can be recursive (using \texttt{let rec}) and get their actual
parameters in a call by tuple-matching their actual argument. 

\subsubsection{Events.\quad} All events come out of the candidate execution and there is no way in \cat\ to generate any other event.  

\subsubsection{Sets} (see Appendix~\ref{sec:sets-tuples-sem} and
Figure~\ref{fig:sets-and-tuples-sem} for their formal semantics) are either
empty \texttt{\{\}} or a homogeneous set \texttt{\{$o_1$, \ldots, $o_n$, \ldots
\}}. We do not allow sets of functions or procedures. Predefined sets of events
are denoted by the following identifiers (see Appendix~\ref{sec:predef-sets}
and  Figure~\ref{fig:predef-sets-sem} for their formal semantics):
\begin{itemize}
\item the set of all write events \catW, including the initial writes \catIW;
\item the set of all read events \catR;
\item the set of all branch events \catB;
\item the set of all fence events \catF;
\item the universe containing all events of the candidate execution, which is denoted ``\verb\_\''.
\end{itemize}

New sets can be defined from existing ones using the following operations (see
Appendix~\ref{sec:operators-sets-relations-sem}, Figures~\ref{fig:ops},
\ref{fig:unary-ops} and~\ref{fig:binary-sem} for the formal semantics of these
operations):
\begin{itemize}
\item the $\catnegation S$ is the complement of a set $S$;
\item the union of two sets $S_1$ and $S_2$  is $S_1 \mid S_2$; 
\item the intersection of two sets $S_1$ and $S_2$  is $S_1$ \texttt{\&} $S_2$; 
\item the difference of two sets $S_1$ and $S_2$  is $S_1 \setminus S_2$ ; 
\item the addition of an element \texttt{e} to a set $S$ is \texttt{e++}$S$;
\end{itemize}

Matching over sets (see Appendix~\ref{sec:sets-tuples-sem} and Figure
\ref{fig:sets-and-tuples-sem} for the formal semantics) can be used for
(recursive) set definitions. Match is against the empty set \texttt{\{\}} or,
for a non-empty set, a partition \texttt{e ++ es} into a singleton
\texttt{\{e\}} and the rest of the set \texttt{es}. 
For example, given a function~$f$, a set~$S=\{e_1, e_2, \dots,e_n\}$ and an
element~$y$, the call \catfold\texttt{\ $f$($S$, $y$)} returns the value
$f(e_{i_1} , f(e_{i_2}, \dots f(e_{i_n},y)))$, where $i_1, i_2, \dots, i_n$ is
some permutation of $1, 2, \dots, n$:
\begin{quote}
\begin{verbatim}
let fold f =
  let rec fold_rec (es,y) = match es with
  || {} -> y
  || e ++ es -> fold_rec (es, f(e,y))
  end
in fold_rec
\end{verbatim}
\end{quote}

\subsubsection{Relations between events}\label{sec:Relation-between-events} (see
Appendix~\ref{sec:predef-rlns} for their formal semantics) can be the empty
relation \texttt{0}, the identity relation \texttt{id}, or the relations
defined from the candidate execution:
\begin{itemize}
\item the program order \texttt{po};
\item the read-from $\catrf$,
\end{itemize}
or predefined relations on events (see Figure \ref{fig:predef-rlns}):
\begin{itemize}
\item the relation \texttt{loc} between events accessing the same memory location;
\item the relation \texttt{ext} between events coming from different threads.\end{itemize}
New relations (see Appendix~\ref{sec:operators-sets-relations-sem}) can be
defined from sets of events (see Figure \ref{fig:binary-sem}):
\begin{itemize}
\item the cartesian product of two sets of events $S_1$ and $S_2$ is $S_1$\texttt{*} $S_2$
\end{itemize}
 or using unary operators on relations (see Figure \ref{fig:unary-ops}):
\begin{itemize}
\item the identity closure of a relation \texttt{r} is \texttt{r?}
\item its reflexive-transitive closure is \texttt{r*}
\item its transitive closure is \texttt{r+}
\item its complement is $\catnegation$\texttt{r}
\item its inverse is \texttt{r\^{}-1}
\end{itemize}
or using binary operators on relations (see Figure \ref{fig:binary-sem}):
\begin{itemize}
\item the union of two relations \texttt{r1}
and \texttt{r2} is \texttt{r1} $\mid$ \texttt{r2} 
\item the intersection of two relations \texttt{r1}
and \texttt{r2} is \texttt{r1  \& r2} 
\item the difference of two relations \texttt{r1}
and \texttt{r2} is \texttt{r1} $\setminus$ \texttt {r2} 
\item the sequence of two relations \texttt{r1} and \texttt{r2} is \texttt{r1;r2} (\ie\ the set
of pairs $(x,y)$ such that there exists an intervening $z$, such that $(x,z)
\in \texttt{r1}$ and $(z,y) \in \texttt{r2}$).
\end{itemize}

Moreover the following primitives can be used to manipulate sets and relations over events (see Figure~\ref{fig:primitives} for their formal semantics): 
\begin{itemize}
\item \catclasses\ takes a relation $r$; if $r$ is an equivalence relation,
then we return the equivalence classes of $r$, otherwise an $\caterror$ is
raised;
\item \catlinearisations\ takes a set $S$ and a relation $r$ and returns a set
of relations; \viz{}, if the relation $r$ is acyclic, we return all the
possible linearisations (topological sorts in the finite case) of $r$ over $S$,
otherwise we return the empty set.  \end{itemize}

\subsubsection{Tuples} (see Appendix~\ref{sec:sets-tuples-sem} and Figure
\ref{fig:sets-and-tuples-sem} for their formal semantics) include the empty
tuple \texttt{()}, and constructed tuples \texttt{($o_1$, \ldots, $o_n$)}.
Tuples can be heterogeneous. Tuples are essentially used to pass parameters to
functions and procedures. Tuples can be destructured by pattern matching; for
example in the example of {\tt fold} above, we match the argument of
\texttt{fold\_rec} into the pair \texttt{(es,y)}.

\subsubsection{Tags.\quad}\label{sec:tags}Events can be tagged (using the
annotations on the program instruction generating this event) and these tags
can be used to build relations. The tags must be declared (see
Appendix~\ref{sec:tag-sem} and Figure \ref{fig:decls}) using the {\tt enum}
construct. For example
\begin{quote}
\begin{verbatim}
enum memory-order = 'rlx || 'acq || 'rel
\end{verbatim}
\end{quote}
defines an enumeration type \texttt{memory-order}, which contains three tags:
\texttt{'rlx} (relaxed), \texttt{'acq} (acquire), \texttt{'rel} (release). 

\LISA instructions can be annotated with such tags\makeatletter\@ifundefined{r@fig:mp-relacq}{}{  (see Figure
\ref{fig:mp-relacq})}\makeatother. In \cat, tags have a quote \texttt{'} to not be confused
with identiers. This confusion is impossible in \LISA so quotes \texttt{'} are
omitted. The tags that can be worn by instructions must be declared (see Figure
\ref{fig:decls} in Appendix \ref{Tags-sem}), as follows:
\begin{quote}
\begin{verbatim}
instructions W[{'rlx,'rel}]
instructions R[{'rlx,'acq}]
\end{verbatim}
\end{quote}

Events generated by an annotated \LISA instruction will bear the same tags as
the instruction.  The set of events bearing a given tag $t$ is provided by
\cattagtoevents \texttt{($t$)} (see Figure \ref{fig:tags} in the Appendix
\ref{Tags-sem}). For example
\begin{quote}
\begin{verbatim}
let Release = tag2events('rel)
let Acquire = tag2events('acq)
\end{verbatim}
\end{quote}
define the set {\tt Release} (resp. {\tt Acquire}) of events bearing the tag
{\tt 'rel} (resp. {\tt 'acq}). 

Tags can be matched against their names as defined in an {\tt enum}, and with
the wildcard \verb+_+ (see Figure \ref{fig:tag-match-sem}); examples are
provided in the next section.

\subsubsection{Scopes.\quad}\label{sec:Scopes}
The organisation of a parallel system is not always flat. Often, threads (and
physical processors or cores alike) are organised in a hierarchical fashion,
threads being members of a hierarchy of nested levels, or \emph{scopes}.
Examples include: the eponymous scope notion in GPU models (\eg{} Cooperative
Thread Array, or {\tt cta}, in Nvidia PTX), or the notion of shareability
domain in ARM (\eg{} {\tt ish} in ARMv8).

Scopes (see Appendix~\ref{sec:scopes-sem} for their formal
semantics\makeatletter\@ifundefined{r@sec:baby-scopes}{}{ and
Section~\ref{sec:baby-scopes} for an example}\makeatother) are special tags
which must be declared with the reserved identifier \texttt{scopes}:
\begin{quote}\small
\begin{verbatim}
enum scopes = 'cta || 'gpu || 'system
\end{verbatim}
\end{quote}
The hierarchy of scopes is described in a \cat\ file by the functions
\texttt{narrower} and  \texttt{wider} (which are reserved identifiers but
user-defined, as \texttt{scopes} is). In the most simple and frequent case,
levels are totally ordered. Then, the \texttt{wider} function takes a scope tag
as argument and returns the immediately wider scope tag, while the
\texttt{narrower} function returns the immediately narrower scope
tag:\footnote{One may also consider heterogeneous systems such as coupled CPUs
and GPUs. In that case, the hierarchical is no longer total and the function
\texttt{narrower} returns a set of tags.}
\begin{quote}
\begin{verbatim}
let wider(s) = match s with 'gpu -> 'system || 'cta -> 'gpu end
let narrower(s) = match s with 'system -> 'gpu || 'gpu -> 'cta end
\end{verbatim}
\end{quote}
The above definitions specify that scopes are ordered from narrowest to widest
as: $\verb+'cta+ < \verb+'gpu+ < \verb+'system+$. In other words, a system
contains one or more GPUs, and each GPU contains one or more CTAs.

All \LISA litmus tests specify how many threads \texttt{P0}, \texttt{P1}, \etc\
are involved. Additionally a scoped litmus test specifies how threads are
distributed along the scope hierarchy, by means of a \emph{scope tree} such as
\bgroup\small
\begin{verbatim}
  scopes: (system (gpu (cta P0 P1) (cta P2 P3)) (gpu (cta P4 P5) (cta P6 P7)))
\end{verbatim}
\egroup%
\noindent which describes the scope hierarchy 
\begin{quote}
\centering\includegraphics[width=0.75\textwidth]{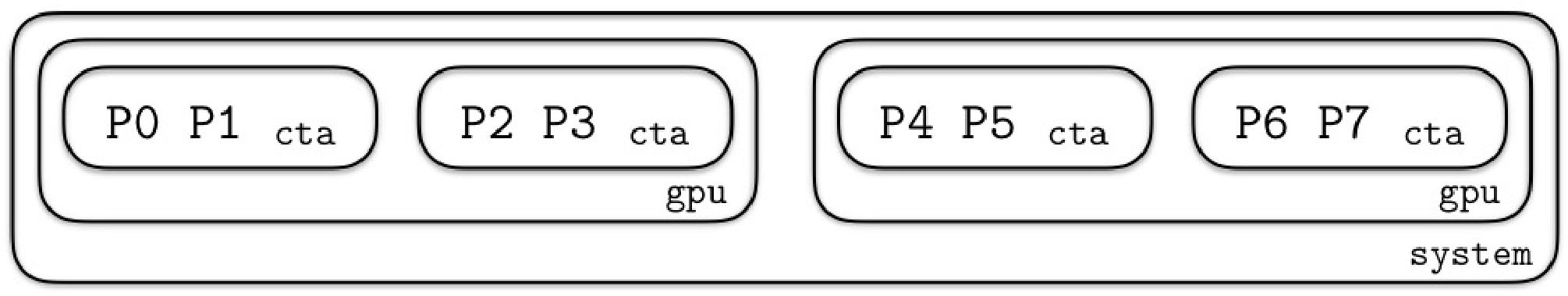}
\end{quote}
The \textsf{herd7} tool checks that the \texttt{wider} function does define a
hierarchy, in the sense that each scope has an unique immediately wider scope
except one, the root of the hierarchy, which has none. It also checks the
compatibility of the \texttt{narrower} function and of scope trees with the
defined hierarchy.

The events in a given scope $s$ are gathered as an equivalence relation
\cattagtoscope\texttt{($s$)} (see Figure \ref{fig:tags} in the Appendix
\ref{sec:scopes-sem}). More precisely two events are related by
\cattagtoscope\texttt{($s$)} when they are generated by threads that are
contained in the same scope instance of level~$s$. 

Consider for instance the hierarchy depicted above, and two events $e_0$, and
$e_2$, generated by \texttt{P0} and \texttt{P2} respectively. Then $e_0$ and
$e_2$ are related by \cattagtoscope\texttt{('gpu)} and unrelated by
\cattagtoscope\texttt{('cta)} since \texttt{P0} and~\texttt{P2} belong to the
same GPU but to different CTAs. \makeatletter\@ifundefined{r@sec:baby-scopes}{}{ We give more examples in Section~\ref{sec:baby-scopes}.}\makeatother

\subsection{Constraint statements}
After defining sets and relations depending on the candidate execution,
we can impose constraints on them (see Figure \ref{fig:checks-sem} for the
formal semantics).

\subsubsection{Checks} (see Appendix \ref{sec:checks} and Figure
\ref{fig:checks-sem} for their formal semantics) can have the following syntax:
$[\catnegation][\,\texttt{acyclic}\,|\,\texttt{irreflexive}\,|\,\texttt{empty}
]\, x$.

The checks $[\catnegation][\,\texttt{acyclic}\,|\,\texttt{irreflexive}\,|]\, r$
check if the relation $r$ on events is be acyclic or irreflexive. The check or
$[\catnegation][\,\texttt{empty} ]\, S$ checks if the set $S$ is empty. The
check $\texttt{acyclic}\,r$ is a shorthand for
$\texttt{irreflexive}\,r\texttt{+}$. The symbol $\catnegation$ denotes the
negation. Failed checks reject the candidate execution which is therefore
\texttt{forbidden}. For example
\begin{quote}
\begin{verbatim}
acyclic po | rf 
\end{verbatim}
\end{quote}
checks whether the union of {\tt po} and {\tt rf} is acyclic in all the
candidate executions of a given program.

Users have the option to not enforce the checks, but rather to use them to
report properties of the candidate execution. To do so, users must prefix the
check they are interested in with the keyword \texttt{flag}, and name the
flagged test with an identifier \textit{name} (by using the postfix qualifier
\texttt{as \textit{name}}). A failed flagged check has no consequence over the
acceptance or rejection of the candidate execution. It is simply reported
(\viz{}, flagged with the name \textit{name}) for the user's information. For example
\begin{quote}
\begin{verbatim}
flag ~(acyclic po | rf) as cycle-found
\end{verbatim}
\end{quote}
will flag, using the name {\tt cycle-found}, all the candidate executions in
which there is a cycle in the union of {\tt po} and {\tt rf}.

We often use {\tt flag} in models that involve \emph{data races}, \eg{} C++ or
HSA. In such models, executions that have data races are typically deemed
undefined. We handle this in \cat{} by flagging candidate executions that
exhibit data races with the name {\tt undefined}.

\subsubsection{Procedures.\quad(see Appendix \ref{sec:procedures-sem} and
Figure \ref{fig:proc-sem} for their formal semantics)}

Definitions of sets and relations and their checks in constraint statements can
be gathered and parameterised using procedures and checked by procedure calls.
Procedures are not recursive and return no result. They have one formal
parameter (but that can be a tuple, including the empty one). Their body is a
non-empty list of statements. 

For example the following procedure \texttt{sc} implements Sequential
Consistency~\cite{lam79}, given the relation {\tt com} as parameter\makeatletter\@ifundefined{r@fig:sc-cat}{}{---we go
back as to how this implements SC in Figure~\ref{fig:sc-cat}}\makeatother:
\begin{quote}
\begin{verbatim}
procedure sc(com) =
   let sc-order = (po | com)+ 
   acyclic sc-order
end
\end{verbatim}
\end{quote}
The procedure may have local definitions (like \texttt{sc-order}). The scope of
the formal parameter and local definitions is limited to the procedure body.
Global definitions (like the relation \texttt{po}) can be used in the procedure
body. 

A call (\eg{} \texttt{call sc(rf)}) passes the actual parameter (a tuple, here
{\tt rf}, matching the formal parameter {\tt com}) and the procedure body is
evaluated with the actual parameter.

\subsubsection{Iteration.\quad}
A universally quantified check (\ie~a finite conjunction of checks) can be done
for all values \texttt{e} chosen in a set \texttt{S} by a \texttt{forall}
iterator (see Appendix \ref{sec:iteration-over-sets} and Figure
\ref{fig:iter-sem} for the formal semantics), as follows:
\begin{quote}
\begin{verbatim}
forall e in S do
   call check_contraint(e)
end
\end{verbatim}
\end{quote}

\subsubsection{Candidate execution extension (\texttt{with ... from ...})\quad}
The construct \texttt{with $o$ from $S$ \catrequirements} introduces an
additional constituant $o$ of the semantics, not already part of the candidate
execution. This constituant $o$ of the semantics is introduced in the \cat file
rather than in the anarchic semantics because it only depends on the program
execution events (\eg{} the coherence order). 

The {\tt with} construct enumerates all possible objects $o$ in $S$ and checks
the \catrequirements. Typically $S$ is a set of relations on events and $o$ a
relation between events which must satisfy the \catrequirements appearing in
the remainder of the \cat{} file. 

For example total orders over certain accesses can be built using {\tt with}:
\begin{itemize}
\item the coherence order between writes to a given memory location 
\makeatletter\@ifundefined{r@sec:building-co}{;}{ (we give
such an example in Section~\ref{sec:building-co});}\makeatother

\item SC accesses in C++ or HSA (we use this in our
modelisation of HSA, see~\cite{HSA-Foundation-PSAS-2015}).
\end{itemize}

\subsection{Evaluation of a \cat\ file on a candidate execution} When evaluated
on a candidate execution, a \cat\ file returns an $\caterror$ if the \cat\ file
is syntactically incorrect. Otherwise the binding definitions, constraint
statements, and \texttt{with} requirements are evaluated in sequence (see
Figures \ref{fig:stat-sem} and \ref{fig:spec-requirements}). If some (unflagged
\ie mandatory) constraint fails, we return $\catforbidden$ to stipulate that
the candidate execution does not satisfy the weak consistency model specified
by the \cat{} file. Otherwise the candidate execution is accepted, \ie{} we
return $\catallowed$. In both cases we return a possibly empty set of flags for
conditions that are not enforceable (see Appendix \ref{sec:specifications}), as
well as the objects introduced by \texttt{with} constructs.

\newpage
\section{Syntax and formal semantics of the
{\ttfamily\bfseries\fontsize{11}{11}\selectfont cat}
language}\label{sec:lisa-cat}

\subsection{Analytic semantics} 

The analytic semantics $\semantics[\texttt{P}]$ of a parallel program $\texttt{P}$ with a given \cat consistency specification (or weak consistency model) \catfile{} is a set of execution behaviors $\execution$ conforming to this consistency specification. Each such  execution behavior $\execution=\triple{\computation}{\catrf}{\communicationrelation}$ is described in two parts, 
the computations $\pair{\computation}{\catrf}$ and the communications $\pair{\catrf}{\communicationrelation}$ where the \emph{read-from relation} $\catrf$  is their common interface.
\begin{itemize}
\item The possible \emph{computations} $\pair{\computation}{\catrf}$ are described by the anarchic semantics $\anarchicsemantics[\texttt{P}]$ of the program $\texttt{P}$. The read-from relation \catrf{} records the
correspondance between the reads, the matching writes, and the communicated values
on the computation $\computation$. 

\mbox{}\hskip1.5em The anarchic semantics $\anarchicsemantics[\texttt{P}]$ places only the following restrictions on the communications of \texttt{P} so all possible computations with all possible read-from relations \catrf{} are considered. 
\begin{itemize}[align=left,leftmargin=*,topsep=0pt,itemsep=0pt,parsep=0pt]

\item \textit{Satisfaction}: a read event has at least one corresponding
communication in $\catrf$;

\item \textit{Singleness}: a read event must have at
most one corresponding communication in $\catrf$;

\item \textit{Match}:  if a read reads from a write, then the variables read
and written and communicated value must be the same;

\item \textit{Inception}: no communication is possible without the occurrence
of both the read and (maybe initial) write it involves (this does not prevent a read to read from a future write).

\end{itemize}
Otherwise stated the consistency specification/weak consistency model is not taken into account at all by the anarchic semantics $\anarchicsemantics[\texttt{P}]$.

\item The possible \emph{communications} are described by communications $\pair{\catrf}{\communicationrelation}$ between communication \ie read and/or write events. 

\noindent\mbox{}\hskip1.5em The \cat file \catfile{} generates all possible communication relations $c\in\communicationrelation$ (using the \catwithtt construct). The communication relations $c\in\communicationrelation$  include the coherence order \texttt{co}, \etc.  More generally, they specify requirements on the execution environment of the program \texttt{P}.

\end{itemize}
The \cat file semantics sorts out the executions $\execution=\triple{\computation}{\catrf}{\communicationrelation}$ that are feasible for weak consistency model, one by one.

\afterpage{\clearpage}
\subsection{Consistent semantics specification by \cat files} 

A \cat file $\catfile\in\setofallcatfiles$ defines a check that an execution $\execution=\triple{\computation}{\catrf}{\communicationrelation}$ satisfies a consistency specification.

\begin{itemize}
\item First the computation $\pair{\computation}{\catrf}$ of the anarchic semantics is abstracted to a candidate execution $\catconfiguration$ = $\defalphacandidateexecution(\pair{\computation}{\catrf})$ = $\quintuple{\setofevents}{\catpo}{\catrf}{\catIW}{\scoperelationitalique}$ (collecting read, write, branch, fence and rmw events in $\setofevents$, the program order $\catpo$, the read-from relation \catrf, initial \catIW writes, and the program scope tree $\scoperelationitalique$) but where \eg  events on local registers or communicated values are abstracted away. 

\item  The \cat file $\catfile$ is then evaluated on $\catconfiguration$. Thanks to \catwithtt $c_i$ \catfromtt $C_i$ constructs, the \cat file generates all necessary communication relations $\communicationrelation$ = $c_1,\ldots,c_n$ between communication events (including \texttt{co} $\in$ \texttt{allCo}, \etc) which are necessary to express the consistency specification.

\item In absence of $\caterror$ in $\catfile$, the final result

\begin{itemize}

\item  can be $\triple{\catallowed}{f}{\communicationrelation}$ meaning that the computation  $\pair{\computation}{\catrf}$ (with abstraction $\catconfiguration$ = $\alphacandidateexecution(\pair{\computation}{\catrf})$) together with the communication specification $\communicationrelation$ satisfies the consistency specification, or

\item can also be $\triple{\catforbidden}{f}{\communicationrelation}$ meaning that the $\execution=\triple{\computation}{\catrf}{\communicationrelation}$ does not satisfy the consistency specification. 

\end{itemize}
In both cases $f\in\catflagdomain$ is the set of flagged constraints in $\catfile$ satisfied by the execution $\execution=\triple{\computation}{\catrf}{\communicationrelation}$ (without any influence on the  \catallowed/\catforbidden \ result).
\end{itemize}

\subsection{Analytic semantics specified by an anarchic semantics and a \protect\cat specification} 

We define below, in Figure \ref{fig:spec-sem}, the semantics $\catsemantics[\catfile]\:\catconfiguration$ of a candidate execution $\catconfiguration$ which returns a set of answers of the form 
$\triple{j}{f}{\communicationrelation}$ where $j=\{\catallowed,\catforbidden\}$, $f$ is the set of flags that have been set up on $\catconfiguration$ and $\communicationrelation$, and $\communicationrelation$ defines the communication relation for the execution to be \catallowed/\catforbidden. 

The analytic semantics of a program \texttt{P} with consistency specification $\catfile$ is therefore
\begin{eqntabular*}{rcl}
\semantics[\texttt{P},\catfile]&\triangleq&\{\triple{\computation}{\catrf}{\communicationrelation}\mid
\begin{array}[t]{@{}l@{}}
\pair{\computation}{\catrf}\in\anarchicsemantics[\texttt{P}]
\wedge{}\\[0.33ex]
\exists f\in\catflagdomain\suchthat
\triple{ \catallowed}{f}{\communicationrelation}\in\catsemantics[\catfile]\:(\alphacandidateexecution(\pair{\computation}{\catrf}))\}
\end{array}
\end{eqntabular*}
This analytic semantics $\semantics[\texttt{P}]$ of a program $\texttt{P}$ for a  \cat\ specification $\catfile$ is the composition $\semantics[\texttt{P}]$ =
$\alphacat[\catfile]\comp\alphacandidateexecution(\anarchicsemantics[\texttt{P}])$ of two abstractions of the anarchic semantics \viz
\begin{eqntabular*}{rcl}
\defalphacandidateexecution(\semantics)
&\triangleq&
\{\pair{\pair{\computation}{\catrf}}{\alphacandidateexecution(\pair{\computation}{\catrf})}\mid
\pair{\computation}{\catrf}\in\semantics\}
\\
\defalphacat[\catfile](\mathcal{C})
&\triangleq&
\{\triple{\computation}{\catrf}{\communicationrelation}\mid
\begin{array}[t]{@{}l@{}}
\pair{\pair{\computation}{\catrf}}{\catconfiguration}\in\mathcal{C}
\wedge{}\\[0.33ex]
\quad\exists f\in\catflagdomain\suchthat
\triple{ \catallowed}{f}{\communicationrelation}\in\catsemantics[\catfile]\:\catconfiguration\}
\end{array}
\end{eqntabular*}

\afterpage{\clearpage}
\subsection{Candidate executions}\label{sec:Candidate-executions-sem}\label{sec:cand-execs}
Candidate executions
are tuples:
\begin{eqntabular}[fl]{@{\quad}rcl}
\catconfiguration&=&\quintuple{\setofevents}{\catpo}{\catrf}{\catIW}{\scoperelationitalique}\colsep{\in}\setofallcatconfigurations\nonumber\\
                   &\triangleq&
\setofallsetsofevents
\times\setofallprogramorders
\times\setofallreadfroms 
\times\setofallsetsofwrites
\times\setofallscoperelations\nonumber
\end{eqntabular}
which gather the events, the program order $\catpo$ on each thread, the read-from relation $\catrf$, modeling who reads from where, the initial writes \catIW, and a scope relation $\scoperelationitalique$.

\subsubsection{Events} $e\in\setofallevents$ are abstractions of the events generated by a program execution. Events $\event\in\setofevents$ carry the unique program instruction (and its unique program label) which execution generated this event $\event$. However, this information is not 
directly available to \cat.  Auxiliaries to extract components of an event $e$ are as
follows:
\begin{eqntabular*}{@{}c@{}}
\begin{array}{@{}rcl@{\quad\quad}rcl@{}}
\deflocationofevent(e)&\triangleq& \text{location of $e$} &
\defcatkindofevent(e)&\triangleq& \text{kind of $e$}
\end{array}\\
\begin{array}{@{}rcl@{\quad\quad}rcl@{}}
\defaidofevent(e)&\triangleq&\text{process identifier of $e$} &
\deftagsofevent(e)&\triangleq&\text{annotations of $e$}
\end{array}\\
\begin{array}{@{}rcl}
\catfromtoofevent(e)&\triangleq&\text{events separated by fence event $e$}
\end{array}
\end{eqntabular*}

The set $\defsetofevents$ of events belong to $\setofallsetsofevents\triangleq\wp(\setofallevents)$.
\begin{itemize}
\item The process identifier $\aidofevent(e)$ refers to the identifier of the unique process at the origin of the event $e$;

\item The location $\locationofevent(e)$ can be a memory location or a register;

\item $\catkindofevent(e)$ is the kind of event $e$: write (\catW), read (\catR), branch (\catB), fence (\catF), begin or end of a \catrmw. We define the following sets of events by kind:
\begin{eqntabular*}{rcl@{\qquad}rcl}
\defsetofwevents(\catconfiguration)&\triangleq&\{\event\in\eventsof(\catconfiguration)\mid\catkindofevent(\event)=\catW\}
&
\defsetofrevents(\catconfiguration)&\triangleq&\{\event\in\eventsof(\catconfiguration)\mid\catkindofevent(\event)=\catR\}
\\
\defsetoffevents(\catconfiguration)&\triangleq&\{\event\in\eventsof(\catconfiguration)\mid\catkindofevent(\event)=\catF\}
&
\defsetofbevents(\catconfiguration)&\triangleq&\{\event\in\eventsof(\catconfiguration)\mid\catkindofevent(\event)=\catB\}
\end{eqntabular*}

\item the annotations $\tagsofevent(e)$ of $e$ is a possibly empty set of tags and scopes carried by the action at the origin of $e$;

\item let $\event_{\!F}\in\setoffevents(\catconfiguration)$ be a fence event generated by a localised fence instruction \texttt{f[\textit{ts}]\ \{L$_1$,\ldots,L$_n$\}\ \{L$'_1$,\ldots,L$'_m$\}} where this instruction and all \texttt{L$_1$, \ldots, L$_n$, L$'_1$, \ldots, L$'_m$} belong to the same process of $\aidofevent(\event_{\!F})$.

Then $\defcatfromtoofevent(\event_{\!F})$ is the set of pairs $\pair{\event_{\!f}}{\event_{\mskip-2mu t}}$ such that $\event_{\!f}$ is an event generated by the execution of a program instruction labelled \texttt{L$_i$}, $i\in\interval{1}{n}$  and $\event_{\mskip-2mu t}$ is an event generated by the execution of a program instruction labelled \texttt{L$'_j$}, $j\in\interval{1}{m}$. Additionally, we require $\pair{\event_{\!f}}{\event_{\!F}}\in \poof(\catconfiguration)$ and $\pair{\event_{\!F}}{\event_{\mskip-2mu t}}\in \poof(\catconfiguration)$, \viz the fence does separate the two events $\event_{\!f}$ and~$\event_{\mskip-2mu t}$. If the fence carries an empty set of labels this is $\catfromtoofevent(\event_{\!F})\triangleq\emptyset$. If the fence carries no sets of labels, we set $\catfromtoofevent(\event_{\!F})\triangleq\{\pair{\event_{\!f}}{\event_{\mskip-2mu t}}\mid\pair{\event_{\!f}}{\event_{\!F}} \in \poof(\catconfiguration)\wedge\pair{\event_{\!F}}{\event_{\mskip-2mu t}} \in \poof(\catconfiguration)\}$.
\end{itemize}

\subsubsection{Program order,} abbreviated  $\defcatpo\in\setofallprogramorders$, abstracts the order of the events of a process in the execution hence lifts the order in which
instructions have been executed to the level of events.
For each candidate execution, it is a total order over events within the
same thread, hence irreflexive and transitive, and cannot relate events from different threads.

\subsubsection{Read-from,} abbreviated $\defcatrf\in\setofallreadfroms\triangleq\wp(\setofallwrites\times\setofallreads)$, relates a read event  of a certain
shared variable \texttt{x} to a unique write event of the same variable. The read-from relation essentially indicates which events read from where.

\subsubsection{Initial writes} are gathered in the set $\defcatIW\in\setofallsetsofwrites\triangleq\wp(\setofallwrites)$.
The initial writes \catIW simply are the writes in the
prelude of the program. 

\subsubsection{Scope relation,}\label{sec:Scope-relation-sem} abbreviated $\defscoperelationitalique\in\setofallscoperelations$, relates
events that come from threads which reside within the same \emph{scope}; this
is a notion that is mostly used for scoped models such as GPUs (see \eg,
\cite{abd15} and Sections~\ref{sec:Scopes} \makeatletter\@ifundefined{r@sec:baby-scopes}{}{ and \ref{sec:baby-scopes}}\makeatother). 

\bigskip

Auxiliaries to extract components of a candidate execution
$\catconfiguration\triangleq\quintuple{\setofevents}{\catpo}{\catrf}{\catIW}{\scoperelationitalique}$ are as follows:
\begin{eqntabular*}{c}
\begin{array}{rcl@{\qquad}rcl@{\qquad}rcl}
\defeventsof(\catconfiguration)&\triangleq&\setofevents 
&
\defpoof(\catconfiguration)&\triangleq&\catpo
&
\defcatrfof(\catconfiguration)&\triangleq&\catrf 
\end{array}
\\
\begin{array}{rcl@{\qquad}rcl}
\defsrof(\catconfiguration)&\triangleq&\scoperelationitalique
&
\definitof(\catconfiguration)&\triangleq&\catIW 
\end{array} 
\end{eqntabular*} 

\subsection{Program scope relation defined by a scope tree and \cat scope hierarchy}\label{sec:Values-sem}

A program scope tree specifies a scope relation. The syntax of program scope trees
and their semantics, that is the scope relation that they define are defined in Figure \ref{fig:scope-trees-sem}. Program scope trees must match the scope hierarchy defined by the \cat file through a scope tag declaration (see Figure \ref{fig:decls}) and the user specified functions with reserved names \catnarrower{} and \catwider{}, as checked in Figure \ref{fig:scopes}.

\begin{figure}[!ht]
\begin{pcframed}[Scope trees]
\begin{eqntabular}[fl]{@{\quad}rcl@{}}
\defst&\grdef&(\stag\ \texttt{P}_0\ \ldots\ \texttt{P}_n)\renumber{program scope trees}\\
   &\gror&(\stag\ \st_0\ \ldots\ \st_n)\nonumber
\end{eqntabular}
where $\{\texttt{P}_0,\ldots,\texttt{P}_n\}\subseteq \{\aidofevent(e)\mid e\in\eventsof(\catconfiguration)\}$. 

\medskip
\mbox{}\hskip1.5em Given a scope tree $\st$, a set $E$ of events and a scope-tag $\st$, define $\srel(\st)\:E\:\stag$ to be the relation between different events that come from threads which reside in the scope $\stag$, as follows
\begin{eqntabular}[fl]{@{\quad}rcl}
\defsrel((\stag\ \texttt{P}_0\ \ldots\ \texttt{P}_n))\:E\:\stag' &\triangleq&\emptyset\renumber{(when $\stag\neq\stag'$)}\\
\srel((\stag\ \texttt{P}_0\ \ldots\ \texttt{P}_n))\:E\:\stag &\triangleq&
\{\pair{e}{e'}\mid 
\begin{array}[t]{@{}l@{}}
e,e'\in E\wedge \exists i,j\in\interval{0}{n}\suchthat\\[0.33ex]
\quad\aidofevent(e)=\texttt{P}_i\wedge\aidofevent(e')=\texttt{P}_j\}
\end{array}\nonumber\\
\srel((\stag\ \st_0\ \ldots\ \st_n))\:E\:\stag' &\triangleq&\bigcup_{i=0}^{n}\:
\srel(\st_i)\:E\:\stag'
\renumber{(when $\stag\neq\stag'$)}\\
\srel((\stag\ \st_0\ \ldots\ \st_n))\:E\:\stag &\triangleq&
\{\pair{e}{e'}\mid
\begin{array}[t]{@{}l@{}}
\displaystyle e,e'\in E\wedge{}\\[0.33ex]
\quad\exists \texttt{P}_i,\texttt{P}_j\in\bigcup_{i=0}^{n}\processes(\st_0\ \ldots\ \st_n)\suchthat\\[0.33ex]
\qquad\aidofevent(e)=\texttt{P}_i\wedge\aidofevent(e')=\texttt{P}_j\}
\end{array}\nonumber\\
\defprocesses((\stag\ \texttt{P}_0\ \ldots\ \texttt{P}_n))&\triangleq&\{\texttt{P}_0,\ldots,\texttt{P}_n\}\nonumber\\
\defprocesses((\stag\ \st_0\ \ldots\ \st_n))&\triangleq&\bigcup_{i=0}^{n}\processes(\st_i)\nonumber\\
\deftagsofscopetree((\stag\ \texttt{P}_0\ \ldots\ \texttt{P}_n))&\triangleq&
\{\stag\}\nonumber\\
\tagsofscopetree((\stag\ \st_0\ \ldots\ \st_n))&\triangleq&
\{\stag\}\cup\bigcup_{i=0}^{n}\tagsofscopetree(\st_i)\nonumber
\end{eqntabular}
If the program has a scope-tree $\st$ then the candidate execution $\catconfiguration$ must have its
scope relation component $\scoperelationitalique=\srof(\catconfiguration)$ 
be such that for all $\stag\in\tagsofscopetree(\st)$,
$\scoperelationitalique(\stag)=\srel\:(\st)\:(\eventsof(\catconfiguration))\:\stag$.
\end{pcframed}
\caption{Semantics of program scope trees~\label{fig:scope-trees-sem}}
\end{figure}

\afterpage{\clearpage}
\subsection{Values}\label{sec:Values-sem}

The \cat{} language is much inspired by OCaml~\cite{OCaml14}, featuring
for example types, immutable bindings, first-class functions and pattern matching.
However, \cat{} is a domain specific language, with important differences
from OCaml: 
\begin{itemize} 
\item base values are specialised; they are: sets of events, relations over
events, first class functions; there are also tags, including scope tags, akin to C~enumerations or
OCaml constant constructors. There are two structured values: sets of values
and tuples of values, see Figure \ref{fig:sem-doms}.  
\item there is a distinction between expressions in Figure \ref{fig:definitions} that evaluate to some value,
statements in Figure \ref{fig:stat-sem}, which introduce new definitions or constraints, and requirements in Figure \ref{fig:spec-requirements}
which introduce new communication relations on the execution environment and constraints on them.
\end{itemize}

We use the following notations: square brackets $[\dots]$
denote optional components, parentheses $(\ldots)$ denote grouping,
$(\ldots)^\ast$ (resp. $(\ldots)^{+}$) denotes zero, one or several (resp. one
or several) repetitions of the enclosed components.

\afterpage{\clearpage}
\subsection{Consistency specifications}\label{sec:specifications}

Consistency specifications (or \cat files/specifications) \catfile{} filter candidate executions and extend them with communication relations.  In other
words, the semantics $\catsemantics[\catfile]\:\catconfiguration$ of a \cat{} specification $\catfile$ is defined with respect to
a candidate execution $\catconfiguration$ and its result extend it to specify requirements on the execution environment.

\subsubsection{Evaluating a $\protect\catfile$ specification} means allowing or forbidding
that candidate execution. More precisely, evaluating a \cat{} file makes a
result object $\catinstructionresultof{j}{f}{\catenvironment}{\communicationrelationidentifiers}$ evolve, where:
\begin{itemize}

\item \emph{Judgements} $j\in\catjudgementdomain\triangleq\{\defcatallowed,\defcatforbidden\}$ can be of two kinds:  \catallowed{} when a candidate
execution passes all the checks imposed by the \cat specification, or
\catforbidden{} when a candidate execution fails on one of the checks of the
\cat specification $\catfile$.

\item \emph{Flagged checks} $f\in\catflagdomain\triangleq\wp(\setofnonterminal{identifier})$  collect identifiers of checks that have been flagged and are recorded to signal certain executions (\eg,
the ones with data races).

\item \emph{Environments} $\catenvironment\in\setofallcatenvironments$  associate identifiers (which belong to the set
$\setofnonterminal{identifier}$) to typed values; more precisely environments
are partial functions from identifiers to values:
\begin{eqntabular*}[fl]{@{\quad}rcl}
\setofallcatenvironments&\triangleq&\setofnonterminal{identifier}\nrightarrow\catexpressionsemanticdomain\ .
\end{eqntabular*} 
During the evaluation of the \cat\ file $\catfile$, the environment $\catenvironment\in\setofallcatenvironments$ gets augmented with new
definitions as evaluation progresses. It evolves also locally when evaluating functions and procedures, according to the static scoping or block-structured visibility rule.

\item \emph{Sets of communication relation identifiers} $\communicationrelationidentifiers$ record the identifiers of communication relations introduced by a \catwithtt requirement.
\begin{eqntabular*}[fl]{@{\quad}rclcl}
\communicationrelationidentifiers&\in&\setofallsetsofcatrequirementidentifiers&\triangleq&\wp(\setofallcatcommunicationrelationidentifiers)
\end{eqntabular*} 
During the evaluation of the \cat\ file $\catfile$, the set $\communicationrelationidentifiers\in\setofallsetsofcatrequirementidentifiers$ of  communication relation identifiers gets augmented with new identifiers introduced by \catwithtt \nonterminal{id} \catfromtt\ldots\ requirements, see Figure \ref{fig:bind-with}. The relation 
$\catenvironment(\nonterminal{id})$ which is the value of such communication relation identifiers \nonterminal{id} is found in the environment $\catenvironment$. The final verdict in Figure \ref{fig:spec-sem} collects this information in the final result of the \catfile{} evaluation.

\item \emph{Results} collect judgements, flagged checks, environments, and communication relation identifiers or raise
$\caterror$ if needed. 
\begin{eqntabular*}[fl]{@{\quad}rclclcl}
\defcatresult
&=&
\defcatinstructionresultof{j}{f}{\catenvironment}{\communicationrelationidentifiers}
&\in&
\defsetofallresults
&\triangleq&
(\catjudgementdomain\times\catflagdomain\times\setofallcatenvironments\times\setofallsetsofcatrequirementidentifiers)\cup\{\caterror\}
\end{eqntabular*} 
A result may be undefined \eg\ when an implementation might not terminate, for example, when evaluating a non-terminating function. The result can also be $\caterror$ when the \cat file is incorrect. The difference is that an implementation of \cat is assumed to signal $\caterror$ but is not required to report undefined results. The final result is collected in the final verdict
$\triple{j}{f}{\prod_{\nonterminal{id}\in\communicationrelationidentifiers}\catenvironment(\nonterminal{id})}$, see Figure \ref{fig:spec-sem}.

\end{itemize}

Initially, the judgement is  \catallowed, the set of flags is empty,
predefined identifiers are implicitly bound to event sets and relations over events as described in Section~\ref{sec:predef} and Figures \ref{fig:predef-sets-sem} and \ref{fig:predef-rlns}, and the set of communication relation identifiers is empty, see Figure \ref{fig:spec-sem}.

\subsubsection{Specifications} (or \cat\ files) are lists of \emph{requirements} preceded by an
identifier, used for documentation purposes. We give the syntax and semantics
of specifications in Figure~\ref{fig:spec-sem}.

The requirements constitutive of the specification are evaluated in sequence,
until one requirement raises $\caterror$ or \catforbidden, or until the end of
the requirement list. In that latter case, the specification accepts the
candidate execution, hence raises  \catallowed. 

\subsubsection{The final verdict} in Figure~\ref{fig:spec-sem} is given at the top-level, gets rid of the environment, and returns the communication relations in $S$  obtained by finding the value of the communication relation identifiers $\nonterminal{id}$ in the environment. If $S$ is empty, we return
$\catforbidden$ (with unmodified flags). If $S$ contains \caterror, the \caterror{} is returned. 

\begin{figure}[!ht]
\begin{eqntabular}[fl]{@{\quad}rcl}
\catfile&\in&\setofallcatfiles\nonumber\\
\defcatfile	&\grdef&	 \textit{identifier} \nonumber\\
&\gror&\textit{identifier}\ \catrequirements\nonumber
\end{eqntabular}
\begin{eqntabular}[fl]{@{\quad}l}
\catsemantics[\textit{identifier}]\:\catconfiguration
\colsep{\triangleq}\{\triple{\catallowed}{\emptyset}{\emptyset}\}\nonumber\\[1em]
%
%
\catsemantics[\textit{identifier}\ \catrequirements]\:\catconfiguration
\colsep{\triangleq}\catverdict(\catsemantics[\catrequirements]\:\catconfiguration\:\quadruple{\catallowed}{\emptyset}{\emptyset}{\emptyset})\nonumber\\[1em]
%
%
%
%
\catsemantics[\catfile]\colsep{\triangleq}
\{\pair{\catconfiguration}{\communicationrelation}\mid
\begin{array}[t]{@{}l@{}}
\catconfiguration\in\setofallcandidateexecutions
\wedge
\communicationrelation\in\setofallcommunicationrelations
\wedge{}\\[0.33ex]
\qquad\exists f\in\catflagdomain\suchthat
\triple{\catallowed}{f}{\communicationrelation}\in\catsemantics[\catfile]\:\catconfiguration\}
\end{array}\nonumber
\end{eqntabular}
\begin{eqntabular}[fl]{@{\quad}l}
%
\defcatverdict\:\emptyset\colsep{\triangleq}
\{\triple{\catforbidden}{\emptyset}{\emptyset}\}\nonumber\\
\catverdict\:S\colsep{\triangleq}\caterror\renumber{when $\caterror\in S$}\\
\catverdict\:{\{\catinstructionresultof{j_i}{f_i}{\catenvironment_i}{\communicationrelationidentifiers_i}\mid i\in\Delta\}}\colsep{\triangleq}
\{\triple{j_i}{f_i}{\prod_{\nonterminal{id}\in\communicationrelationidentifiers_i}\catenvironment_i(\nonterminal{id})}
\mid i\in\Delta\}
\renumber{otherwise}
\end{eqntabular}
\caption{Semantics of specifications~\label{fig:spec-sem}}
\end{figure}

\subsection{Statements}\label{sec:statements}

Requirements can be \nonterminal{statements} introducing new binding definitions and checking constraints, or the \catwithtt \nonterminal{id} \catfromtt \nonterminal{S} requirement introducing a new communication relation identified by \nonterminal{id}.

\emph{Statements} are evaluated for their effect: adding new \emph{definitions} or checking \emph{constraints}. We give
their syntax and semantics in Figure~\ref{fig:stat-sem}. Note that once an
$\caterror$ has been raised, we stay in that state. Moreover statements have no \catwithtt requirement so cannot introduce new communication identifiers. Therefore the set of communication relation identifiers is unchanged by the evaluation of a statement, $\communicationrelationidentifiers'=\communicationrelationidentifiers$ in Figure~\ref{fig:stat-sem}.
\begin{figure}[!ht]
\abovedisplayskip0.5\abovedisplayskip
\belowdisplayskip0.5\belowdisplayskip
\begin{eqntabular*}[fl]{@{\quad}rcl}
\defnonterminal{statements}&\in&\setofnonterminal{statements}\\
\nonterminal{statements}&\grdef&\{  \catinstruction \}^+\\[0.5ex]
\defcatinstruction &\in&\catsetofallinstructions\\ 
\catinstruction 	&\grdef& \catdefinition\\ 
&\gror&	 \catconstraint
\end{eqntabular*}
\begin{eqntabular*}[fl]{@{\quad}l} 
\catsemantics[\nonterminal{statements}]\colsep{\in}\setofallcatconfigurations\rightarrow\setofallresults\rightarrow\setofallresults\nonumber\\
\catsemantics[\catinstruction\ \nonterminal{statements}]\:\catconfiguration\:\catinstructionresultof{j}{f}{\catenvironment}{\communicationrelationidentifiers}
\colsep{\triangleq}\nonumber\\
\quad\llet\ \catinstructionresultof{j'}{f'}{\catenvironment'}{\communicationrelationidentifiers'}= \catsemantics[\catinstruction]\:\catconfiguration\:\catinstructionresultof{j}{f}{\catenvironment}{\communicationrelationidentifiers}\ \lin\nonumber\\
\quad\quad\iif\ j'=\catallowed\ \ithen\nonumber\\
\quad\quad\quad\catsemantics[\nonterminal{statements}]\:\catconfiguration\:\catinstructionresultof{j'}{f'}{\catenvironment'}{\communicationrelationidentifiers'}\nonumber\\
\quad\quad\ielse\ \catinstructionresultof{j'}{f'}{\catenvironment'}{\communicationrelationidentifiers}
\end{eqntabular*}
\begin{eqntabular*}[fl]{@{\quad}rcl}
\catsemantics[\nonterminal{statement}]\:\catconfiguration\:\caterror
\colsep{\triangleq}
\caterror\nonumber
\end{eqntabular*}
\caption{Semantics of statements~\label{fig:stat-sem}}
\end{figure}

\subsection{Typed values and semantic domains}\label{sec:types-semantic-domains}

\paragraph{Typed values,} (gathered in the set $\catexpressionsemanticdomain$)
are given in Figure~\ref{fig:types}. Events (of type $\sffont{evt}$) belong to
the set $\setofnonterminal{evt}$.  There are no operation on events so the type
$\sffont{evt}$ can only be used to type elements of relations or sets. Typed
values include (see Figure \ref{fig:sem-doms}):
\begin{itemize}
\item the $\caterror$ symbol;
\item tags (of type $\sffont{tag}$), which belong to $\setofnonterminal{tag}$;
\item relations over events (of type $\sffont{rel}$), which belong to
$\wp(\setofnonterminal{evt}\times\setofnonterminal{evt})$;
\item sets (of type $\sffont{set}$) of values, which belong to
$\wp(\catexpressionsemanticdomain)$; sets have to be homogeneous, and cannot be
sets of functions or procedures, as reflected by the predicate
$\sffont{well-formed}$;
\item tuples (of type $\sffont{tuple}$) of values, which belong to
$\bigcup_{n\in\setofallnaturals}\prod_{i=1}^{n}\catexpressionsemanticdomain$;
\item enumerations of tags (of type $\sffont{enum}$), which belong to
$\wp(\setofnonterminal{tag})$;
\item functions (of type $\sffont{fun}$);
\item non-recursive procedures (of type $\sffont{proc}$).
\end{itemize}

The value of functions and procedures are closures memorising their parameter
(which belongs to \nonterminal{Pat}), their body (which in the case of
functions belongs to \nonterminal{Expr}, and in the case of procedures can be a
list of elements of \nonterminal{Statement}), and declaration \emph{environment} (which belongs to $\setofallcatenvironments$). On a call, the actual parameters are evaluated in the calling environment and the body in the declaration environment enriched by the value of the formal parameters and the local bindings. After the call, evaluation goes on in the calling environment. This is therefore static scoping.

\newcommand{\defwellformedset}[1]{\hypertarget{hyper:wellformedset}{\mathsffont{well-formed}(#1)}}
\newcommand{\wellformedset}[1]{\hyperlink{hyper:wellformedset}{\mathsffont{well-formed}(#1)}}
\begin{figure}[!ht] 
\begin{eqntabular}[fl]{@{\quad}rcl}
\defcattypenocolon{type}{}&\in&\textit{Type}\nonumber\\
\cattypenocolon{type}{}&\grdef&\nonumber\\
&&\defcattypenocolon{evt}{}\renumber{events}\\
&\gror&\cattypenocolon{tag}{}\renumber{tag}\\
&\gror&\cattypenocolon{rel}{}\renumber{relation between events}\\
&\gror&\cattypenocolon{set}{}\renumber{set}\\
&\gror&\cattypenocolon{tuple}{}\renumber{tuple}\\
&\gror&\cattypenocolon{enum}{}\renumber{enumeration}\\
&\gror&\cattypenocolon{fun}{}\renumber{unary function type}\\
&\gror&\cattypenocolon{proc}{}\renumber{unary procedure type} 
\end{eqntabular}
\caption{Typed values \label{fig:types}}
\end{figure}
\begin{figure}[!ht] 
\bgroup
\abovedisplayskip-10pt
\belowdisplayskip-10pt
\begin{eqntabular}[fl]{@{}rcl}
\defwellformedset{S}&\triangleq&\forall\mskip3mu \cattype{type}v\in S\suchthat\cattypenocolon{type}{}\not\in\{\cattypenocolon{fun}{},\cattypenocolon{proc}{}\}
\wedge\forall\mskip3mu \cattypenocolon{type}{$'${:}}v'\in S\suchthat
\cattypenocolon{type}{}=\cattypenocolon{type}{$'$}{}\nonumber\\
\renumber{check sets}
\end{eqntabular}
\egroup
\begin{pcframed}[Semantic domains]
\begin{eqntabular}[fl]{@{\quad}rcl}
\defcatexpressionsemanticdomain&\triangleq&\renumber{typed values}\\ &&
\{\defcaterror\}\nonumber \\ &&
{}\cup{}\defcattypetag\setofnonterminal{tag}\nonumber \\ &&
{}\cup{} \defcattyperel\wp(\setofallevents\times\setofallevents) \nonumber\\&&
{}\cup{} \defcattypeset\{S\in\wp(\catexpressionsemanticdomain)\mid
\wellformedset{S}\}
\nonumber\\&&
{}\cup{}
\defcattypetuple(\{()\}\cup \bigcup_{n\in\setofallnaturals, n>1}\:\prod_{i=2}^{n}\catexpressionsemanticdomain)
\nonumber\\&&
{}\cup{}\defcattype{enum}\wp(\setofnonterminal{tag}) \nonumber\\&&
%
{}\cup{}
\defcattypefun((\setofnonterminal{pat}\rightarrow\setofnonterminal{expr})\times\setofallcatenvironments)\nonumber\\&&
{}\cup{}
\defcattypeproc((\setofnonterminal{pat}\rightarrow\{\catsetofallinstructions\}^+)\times\setofallcatenvironments)
\nonumber\\
\catenvironment\colsep{\in}\defsetofallcatenvironments&\triangleq&\setofnonterminal{identifier}\nrightarrow\catexpressionsemanticdomain\renumber{environments}\\ 
j\colsep{\in}\defcatjudgementdomain&\triangleq&\{\catallowed,\catforbidden\}\renumber{judgements}\\
%
f\colsep{\in}\defcatflagdomain&\triangleq&\wp(\setofnonterminal{identifier})\renumber{flagged checks}\\
\communicationrelationidentifiers\colsep{\in}\defsetofallsetsofcatrequirementidentifiers&\triangleq&\wp(\setofallcatcommunicationrelationidentifiers)\renumber{set of com.\ identifiers}\\
%
%
\catresult\colsep{\in}\defcatinstructionsemanticdomain&\triangleq&
(\catjudgementdomain\times\catflagdomain\times\setofallcatenvironments{\times \setofallsetsofcatrequirementidentifiers}){}\cup{}
\{\caterror\}\renumber{results}
\end{eqntabular}
\end{pcframed}
\caption{Semantic domains\label{fig:types}\label{fig:sem-doms}}
\end{figure}

\subsection{Auxiliaries}
To define the semantics of operators over sets and relations in particular we
need to define a certain number of auxiliaries (summarised in
Figure~\ref{fig:aux-sem}).
\begin{figure}[!ht]
\begin{eqntabular}[fl]{@{\quad}l}
\defidentityrelation_{\mathcal{X}}
\colsep{\triangleq}\{\pair{e}{e}\mid e\in\mathcal{X}\}\renumber{identity relation on set $\mathcal{X}$}\\
r\defrelationcomposition r'
\colsep{\triangleq}\{\pair{\event}{\event'}\mid\exists\event''\suchthat\pair{\event}{\event''}\in r\wedge\pair{\event''}{\event'}\in r'\}\renumber{sequence of relations}\\
\defdomainof{r}\colsep{\triangleq}\{x\mid \exists y\suchthat\pair{x}{y}\in r\}\renumber{domain of relation $r$}\\
\defcodomainof{r}\colsep{\triangleq}\{y\mid \exists x\suchthat\pair{x}{y}\in r\}\renumber{range of relation $r$}\\
\deffieldof{r}\colsep{\triangleq}\domainof{r}\cup\codomainof{r}\renumber{field of relation $r$}\\
\defLfp{\subseteq} F=\bigcap\{X\in \wp(S)\mid F(X)\subseteq X\}\renumber{\parbox[t]{7cm}{the least fixpoint of the $\subseteq$-increasing operator $F$ on the powerset $\wp(S)$~\cite{Tarski55-fp}}}
\end{eqntabular}
\caption{Auxiliaries for defining operators' semantics~\label{fig:aux-sem}}
\end{figure}

\subsection{Expressions}\label{sec:expressions-sem}

Expressions let the user build new sets or relations over tags and events.
Figure~\ref{fig:exprs} summarises the syntax of expressions.

\begin{figure}[!ht]
\begin{eqntabular}[fl]{@{\quad}rcl}
\nonterminal{simple}&\in&\setofnonterminal{Simples}\nonumber\\
\defnonterminal{simple}&\grdef& \nonumber\\	
 	&\gror&	 \nonterminal{id}  \renumber{identifiers}\\
 	&\gror&	 \mtag   \renumber{tags}\\
	&\gror&	 \nonterminal{function} \renumber{anonymous functions}\\
	&\gror&	 \nonterminal{procedure} \renumber{procedures}\\
 &\gror&	  \catsetnonterminal \renumber{sets}\\
	&\gror&	 \cattuplenonterminal  \renumber{tuples}
\end{eqntabular}
%

\begin{eqntabular}[fl]{@{\quad}rcl}
\nonterminal{clause} &\in& \setofnonterminal{clauses} \nonumber \\
\defnonterminal{clause} &\grdef& \nonumber\\
	&&	 [ \catalternative ]\  \mtag \  \mathbin{\texttt{->}}\  \catexpr{}\  \{ \catalternative\  \mtag \  \mathbin{\texttt{->}}\    \catexpr{}\ \}^\ast\  [ \mdefault\  \mathbin{\texttt{->}}\  \catexpr{}] \nonumber \\
&\gror&	 [ \catalternative ]\  \mathopen{\texttt{\{}} \mathclose{\texttt{\}}}\  \mathbin{\texttt{->}}\  \catexpr{}\  \catalternative  \ \nonterminal{id}  \mathbin{\texttt{++}}  \nonterminal{id} \ \mathbin{\texttt{->}} \ \catexpr{}\nonumber
\end{eqntabular}

\begin{eqntabular}[fl]{@{\quad}rcl}
\defcatexpr&\in&\setofnonterminal{expr}\nonumber\\
\catexpr{}&\grdef& \nonumber\\	
        &\gror&\nonterminal{simple} \renumber{simples}\\ 
 	&\gror&	 \catexpr{}\  \catexpr{}  \renumber{function application}\\
 	&\gror&	 \mathopen{\texttt{(}}\catexpr{}\mathclose{\texttt{)}} \gror   
	 \mathopen{\texttt{begin}}\  \catexpr{}\  \mathclose{\texttt{end}}  \renumber{grouping}\\
 	&\gror&	 \texttt{let}\  [\texttt{rec} ] \ \catbinding\  \{ \texttt{and}\  \catbinding \}^\ast \ \texttt{in}\  \catexpr{}  \renumber{binding expressions}\\
  	&\gror&	 \catmatchtt\  \catexpr{}\  \catwithtt\  \nonterminal{clause}\  \texttt{end}  \renumber{matching}\\
       &\gror&\catop \renumber{operators on sets and relations}
%
\end{eqntabular}
\begin{eqntabular*}[fl]{@{\quad}rcl} 
\defcatdefinition &\in&\catsetofalldefinitions\\ 
\catdefinition &\grdef& \nonterminal{decl}\\ 
&\gror&	 \texttt{let} \ [ \texttt{rec} ]\
\catbinding \ \{ \texttt{and}\  \catbinding \}^\ast  
%
\end{eqntabular*}
\caption{Simple expressions, expressions and definitions~\label{fig:exprs}\label{fig:definitions}}
\end{figure}

Several constructs are non-deterministic: the set matching of
Section~\ref{sec:set-match}, the iteration over sets of
Section~\ref{sec:iteration-over-sets}. In the semantics, only one result is nondeterministically picked out of all possible ones. This is different from the \catwithtt requirement of Section~\ref{sec:candidate-execution-extension} where all possibilities for choosing the communication relation are enumerated.

The semantics of an expression is $\caterror$ whenever the semantics of any one
of its subexpressions is $\caterror$. To leave this check implicit, we assume
that the mathematical construct $\llet\ \cattypenocolon{type}{$_i${:}}v_i\ =\
\catsemantics[\mskip3mu\catexpr{}_i]\:\catconfiguration\:\catenvironment,\quad
i\in\interval{1}{\ell}\ \lin\ \ldots$ equals $\caterror$ whenever there exists
$i$ in $\interval{1}{\ell}$ such that
$\catsemantics[\mskip3mu\catexpr{}_i]=\caterror$.


\subsubsection{Identifiers}\label{sec:predef} are either predefined or defined
by the user through \emph{definition} statements. We list the reserved
identifiers in Figure~\ref{fig:reserved-ids}. User-defined identifiers cannot
be reserved identifiers and are bound in the environment $\catenvironment$ (see
Figure \ref{fig:Semantics-of-user-defined-identifiers}).

\begin{figure}[!ht]
\vspace*{-2mm}
\bgroup\newdimen\largeur\largeur\textwidth\advance\largeur by -1.5em
\begin{eqntabular*}[fl]{@{}p{1.5em}@{}p{\largeur}@{}}
\rlap{\sffont{Keywords} $\triangleq$}\\
&
\{\catacyclictt,
\texttt{and},
\texttt{as},
\texttt{begin},
\catcalltt,
\catdott,
\catemptytt,
\texttt{end},
\catenum{},
\catflagtt,
\catforalltt,
\catfromtt,
\catfuntt,
\texttt{in},
\catinstructions,
\catirreflexivett,
\texttt{let},
\catmatchtt,
\catprocedurett,
\texttt{rec},
\catscopes,
\catwithtt\}
\\
\rlap{\sffont{Primitives} $\triangleq$}\\
&
\{\catclasses, \catfromto, \catlinearisations, \cattagtoevents, \texttt{tag2scopes}\}
\\
\rlap{\sffont{Names} $\triangleq$}\\
&
\{$\mdefault$,
\catemptyrelation,
\catB,
\catext,
\catF,
\catid,
\catIW,
\catloc,
\catM,
\catnarrower,
\catpo,
\catR,
\catrf,
\catrmw,
\catW,
\catwider\}\\
\rlap{$\protect\defsetofallreservedidentifiers\colsep{\triangleq} \sffont{Keywords} \cup \sffont{Primitives} \cup \sffont{Names}$}
\end{eqntabular*}%
\egroup

\vspace*{-2mm}
\caption{List of reserved identifiers} \label{fig:reserved-ids}
\end{figure}

\begin{figure}[!ht]
\vspace*{-2mm}
\belowdisplayskip0pt
\begin{eqntabular}[fl]{@{\quad}l}
\nonterminal{id}\in\setofnonterminal{identifier}\nonumber\\
\nonterminal{id}\in\defsetofallcatcommunicationrelationidentifiers\colsep{\triangleq}
\setofnonterminal{identifier}\setminus\setofallreservedidentifiers\nonumber
\end{eqntabular}
\abovedisplayskip0pt\belowdisplayskip1ex
\begin{eqntabular}[fl]{@{\quad}l}
\catsemantics[\nonterminal{id}\mskip2mu]\:\catconfiguration\:\catenvironment
\colsep{\triangleq}\begin{array}[t]{@{}l}
\iif\ \nonterminal{id}\in\domainof{\catenvironment}\ \ithen\\
\quad\llet\ \cattype{type}v=\catenvironment(\nonterminal{id})\ \lin\\
\quad\quad\iif \cattypenocolon{type}{}=\cattypenocolon{enum}{}\ \ithen\ \cattype{set}v\ \:\ielse\ \cattype{type}v\\
\ielse\ \caterror 
\end{array}
\renumber{(when $\nonterminal{id}\not\in\sffont{Names}$)}
\end{eqntabular}
\quad(for $\nonterminal{id}\in\sffont{Names}$, see Figures \ref{fig:predef-sets} or \ref{fig:predef-rlns-sem})
\caption{Semantics of identifiers}\label{fig:Semantics-of-user-defined-identifiers}\label{fig:definition-sem}
\end{figure}

\paragraph{Predefined identifiers denoting sets of
events}\label{sec:predef-sets} appear in Figure~\ref{fig:predef-sets}. We have:
the universal sets, the set of all write, read, memory, branch
and fences events, as well as the set of initial writes. 
The semantics of these identifiers, given in Figure~\ref{fig:predef-sets-sem}
is straightforward; they denote the eponymous sets of events.

\begin{figure}[!ht]
\vspace*{-2mm}
\begin{eqntabular}[fl]{@{\quad}rcl}\label{fig:annot-evts}
\defnonterminal{aevt} &\grdef& \renumber{annotable events}\\
&\gror& \defcatW\renumber{write events\quad}\\
&\gror& \defcatR\renumber{read events\quad}\\
&\gror& \defcatB\renumber{branch events\quad}\\
&\gror& \defcatF\renumber{fence events\quad}
\end{eqntabular}

\begin{eqntabular}[fl]{@{\quad}rcl}
\textit{predefined-events}&\grdef& \nonumber\\
%
%
&\gror& \mdefault\renumber{all events\quad}\\
&\gror& \catIW\renumber{initial writes\quad}\\
%
%
&\gror& \catM\renumber{memory events, $\catM = \catW \cup \catR$\quad}\\
&\gror& \nonterminal{aevt}\renumber{annotable events\quad}
\end{eqntabular}

\begin{eqntabular}[fl]{@{\hskip-1mm}rcl}
%
\catsemantics[{\textbf{\texttt{\_$\mskip-2mu$\_}}}\mskip2mu]\:\catconfiguration\:\catenvironment&\triangleq&\cattypeset\{\cattypeevt \event\mid \event\in\eventsof(\catconfiguration)\}\renumber{events}\\
%
\catsemantics[\catW\mskip2mu]\:\catconfiguration\:\catenvironment&\triangleq&\cattypeset\{\cattypeevt \event\mid \event\in
\eventsof(\catconfiguration)\cap\setofwevents(\catconfiguration)\}\renumber{write events}\\
\catsemantics[\catR\mskip2mu]\:\catconfiguration\:\catenvironment&\triangleq&\cattypeset\{\cattypeevt \event\mid \event\in\eventsof(\catconfiguration)\cap\setofrevents(\catconfiguration)\}\renumber{read events}\\
\catsemantics[\catB\mskip2mu]\:\catconfiguration\:\catenvironment&\triangleq&\cattypeset\{\cattypeevt \event\mid\event\in\eventsof(\catconfiguration)\cap\setofbevents(\catconfiguration)\}
\renumber{branch events}\\
\catsemantics[\catF\mskip2mu]\:\catconfiguration\:\catenvironment&\triangleq&\cattypeset\{\cattypeevt \event\mid\event\in\eventsof(\catconfiguration)\cap\setoffevents(\catconfiguration)\}\renumber{fence events} \\
\catsemantics[\catM\mskip2mu]\:\catconfiguration\:\catenvironment&\triangleq&\cattypeset\{\cattypeevt \event\mid \event\in
\eventsof(\catconfiguration)\cap(\setofrevents(\catconfiguration)\cup\setofwevents(\catconfiguration))\}\renumber{memory events}\\
\catsemantics[\catIW\mskip2mu]\:\catconfiguration\:\catenvironment&\triangleq&\cattypeset\{\cattypeevt \event\mid \event\in
\initof(\catconfiguration)\}\renumber{initial write events}
%
%
%
%
\end{eqntabular}
where $\catsemantics[\catIW\mskip2mu]\:\catconfiguration\:\catenvironment\subseteq\catsemantics[\catW\mskip2mu]\:\catconfiguration\:\catenvironment$  
\caption{Predefined sets and their semantics\label{fig:predef-sets} \label{fig:predef-sets-sem}}
\end{figure}

\afterpage{\clearpage}
\paragraph{Predefined identifiers denoting relations on
events}\label{sec:predef-rlns} appear in Figure~\ref{fig:predef-rlns}. We have:
the empty and identity relations, the relation over events accessing the same
memory location, the relation over events with different pids, the program
order, and the read-from relation.

The semantics of these predefined identifiers, given in
Figure~\ref{fig:predef-rlns-sem} is relatively straightforward again:
\catemptyrelation{} is the empty relation, \catid{} is the identity relation,
\catloc{} the relation between events accessing the same variable, and
\catext{} the relation between events from different threads. It is the
eponymous relation for \catpo and \catrf. 

\begin{figure}[!ht]
\begin{pcframed}[Predefined relations over events]
\begin{eqntabular}[fl]{@{\quad}rcl}
\textit{predefined-relations}&\grdef&\catemptyrelation\renumber{empty relation}\\
&\gror&\catid\renumber{identity}\\
&\gror& \catloc\renumber{same location}\\
&\gror& \catext\renumber{external (different pids)}\\
%
%
&\gror& \catpo\renumber{program order}\\
&\gror& \catrf\renumber{read-from} \\
&\gror& \catrmw\renumber{read-modify-write} 
%
%
\end{eqntabular}
\end{pcframed}

\begin{pcframed}[Semantics of predefined relations]
\begin{eqntabular}[fl]{@{\quad}rcl}
\catsemantics[\defcatemptyrelation]\:\catconfiguration\:\catenvironment&\triangleq&\cattyperel\emptyset\nonumber\\
\catsemantics[\defcatid]\:\catconfiguration\:\catenvironment&\triangleq&\cattyperel\identityrelation_{\eventsof(\catconfiguration)}
\colsep{\triangleq}
\cattyperel\{\pair{\event}{\event}\mid \event\in\eventsof(\catconfiguration)\}\nonumber\\
\catsemantics[\defcatloc]\:\catconfiguration\:\catenvironment&\triangleq&
\cattyperel\{\pair{\event}{\event'}\in\eventsof(\catconfiguration)\times\eventsof(\catconfiguration)\mid\locationofevent(\event)=\locationofevent(\event')\}\nonumber\\
\catsemantics[\defcatext]\:\catconfiguration\:\catenvironment&\triangleq&
\cattyperel\{\pair{\event}{\event'}\in\eventsof(\catconfiguration)\times\eventsof(\catconfiguration)\mid\wedge\aidofevent(\event)\neq\aidofevent(\event')\}\nonumber\\
\catsemantics[\catpo]\:\catconfiguration\:\catenvironment&\triangleq&\cattyperel\poof(\catconfiguration)\nonumber\\
\catsemantics[\catrf]\:\catconfiguration\:\catenvironment&\triangleq&\cattyperel\catrfof(\catconfiguration)\nonumber
%
%
%
\\
%
%
\catsemantics[\defcatrmw]\:\catconfiguration\:\catenvironment&\triangleq&\llet\ RMW\ =\ 
\{\pair{r}{w}\mid \exists \event_b,\event_e\in\eventsof(\catconfiguration)\suchthat{}\nonumber\\
&&
\qquad\catkindofevent(\event_b)=\catbeginrmw\wedge\catkindofevent(\event_e)=\catendrmw\wedge{}\nonumber\\
&&
\qquad \pair{\event_b}{r}\in\poof(\catconfiguration)\wedge{}(\nexists\event\in\eventsof(\catconfiguration)\suchthat\pair{\event_b}{\event}\in\poof(\catconfiguration)\wedge{}\nonumber\\
&&
\qquad\pair{\event}{r}\in\poof(\catconfiguration))\wedge
\pair{r}{w}\in\poof(\catconfiguration){}\wedge\pair{w}{\event_e}\in\poof(\catconfiguration)\wedge{}\nonumber\\
&&
\qquad(\nexists\event\in\eventsof(\catconfiguration)\suchthat\pair{w}{\event}\in\poof(\catconfiguration)\wedge\pair{\event}{{\event_e}}\in\poof(\catconfiguration))\wedge{}\nonumber\\
&&
\qquad
(\nexists\event\in\eventsof(\catconfiguration)\suchthat
(\pair{\event_b}{\event}\in\poof(\catconfiguration)
\wedge \pair{\event}{\event_e}\in\poof(\catconfiguration))\wedge{}\nonumber\\
&&
\qquad\qquad\catkindofevent(\event)\in\{\catbeginrmw,\catendrmw\})\}\nonumber\\
&&
\lin\nonumber\\
&&
\quad\cattyperel RMW\nonumber
\end{eqntabular} 
\end{pcframed} \caption{Predefined relations over events and
their semantics \label{fig:predef-rlns}\label{fig:predef-rlns-sem}}
\end{figure} 

\clearpage
\subsection{Primitives to manipulate sets and relations over
events}\label{sec:primitives} appear in Figure~\ref{fig:primitives}.  We have
five primitives ($\sffont{Primitives}\triangleq\{\catclasses$, $\catfromto$,
 $\catlinearisations$, $\cattagtoevents$, $\cattagtoscope{}\})$.  We will
detail the primitives \cattagtoevents\ and \cattagtoscope\ in
Section~\ref{sec:tag-sem}. For the other three primitives:
\begin{itemize}
\item $\catclasses$ takes as argument an expression \catexpr{}, which
should evaluate as a relation $r$; if $r$ is an equivalence relation, then we
return the equivalence classes of $r$, otherwise we raise an $\caterror$;
\item $\catlinearisations$ takes as argument a pair of two expressions
\catexpr{}$_1$, which should evaluate to a set $S$, and
\catexpr{}$_2$, which should evaluate to a relation $r$; if this
relation is acyclic, then we return all the possible linearisations
(topological sorts) of $r$ over $S$, otherwise we return the empty set.
\item $\catfromto$ takes as argument a expression \catexpr{}, which should evaluate to a set $S$ of tags, the events tagged with these tags should be fence events, and the result is the union of all their sets of pairs of events separated by these fence events.
\end{itemize}
\begin{figure}[!ht]
\begin{pcframed}[Semantics of primitive functions]
\begin{eqntabular}[fl]{@{\quad}l}
%
\catsemantics[\defcatclasses\ \catexpr{}]\:\catconfiguration\:\catenvironment\colsep{\triangleq}
\llet\ \cattype{type}r\ =\ \catsemantics[\catexpr{}]\:\catconfiguration\:\catenvironment\ \lin\nonumber\\
\quad\iif(\cattypenocolon{type}{}=\cattypenocolon{rel}{})\wedge(\identityrelation_{\fieldof{r}}\subseteq r{}\wedge{}
(r)^{-1}\subseteq r\wedge r\relationcomposition r\subseteq r{})\ithen\nonumber\\
\quad\quad\cattypeset\{\cattypeset\{\cattypeevt \evt\in \fieldof{r}\mid \pair{\evt}{\evt'} \in r\}\mid \evt'\in \fieldof{r}\}\nonumber\\
\quad\ielse \caterror
\nonumber\\[1ex]
%
%
\catsemantics[\defcatlinearisations\ \catexpr{}]\:\catconfiguration\:\catenvironment
\colsep{\triangleq}
\llet\ \cattype{type}v \ =\ \catsemantics[\catexpr{}]\:\catconfiguration\:\catenvironment \lin\nonumber\\
\quad\iif\ (\cattype{type}v=\cattypetuple\pair{\cattypeset s}{\cattyperel r})\wedge(\forall\mskip3mu\cattypenocolon{type}{$_v${:}}v\in s\suchthat \cattypenocolon{type}{$_v$}{}=\cattypenocolon{evt}{})\ithen\nonumber\\
%
%
\quad\quad\iif r^+\cap\identityrelation_{s}=\emptyset \ithen\nonumber\\
\quad\quad\quad\cattypeset\{\cattyperel r'\in\wp(s\times s) \mid 
\begin{array}[t]{@{}l@{}} 
r\cap(s\times s)\subseteq{r'}\wedge{}
%
r'\relationcomposition r'\subseteq r'{}\wedge{}\\ 
(\forall\event\neq\event'\in s:\pair{\event}{\event'}\in r'\vee\pair{\event'}{\event}\in r')\}
\end{array}
\nonumber\\
\quad\quad\ielse \cattypeset\emptyset\nonumber\\
\quad\ielse \caterror
\nonumber\\[1ex]
%
%
\catsemantics[\defcatfromto{}\ \catexpr{}]\:\catconfiguration\:\catenvironment
\colsep{\triangleq}\llet\ \cattype{type}S \ =\ \catsemantics[\catexpr{}]\:\catconfiguration\:\catenvironment \lin\nonumber\\
\quad\iif\ (\cattypenocolon{type}{}=\cattypenocolon{set}{})\wedge(\forall\mskip3mu\cattypenocolon{type}{$_e${:}}e\in S\suchthat \cattypenocolon{type}{$_e$}{}=\cattypenocolon{evt}{}\wedge e\in\setoffevents(\catconfiguration))\ithen\nonumber\\
\quad\quad\cattyperel\bigcup_{e\in S}\catfromtoofevent(e)\nonumber\\
\quad\ielse \caterror
\nonumber
\end{eqntabular}
\end{pcframed}
\vskip-3mm
\caption{Semantics of primitives} \label{fig:primitives}
\end{figure}

\afterpage{\clearpage}
\subsubsection{Tags}\label{sec:tag-sem}

\begin{figure}[!ht]
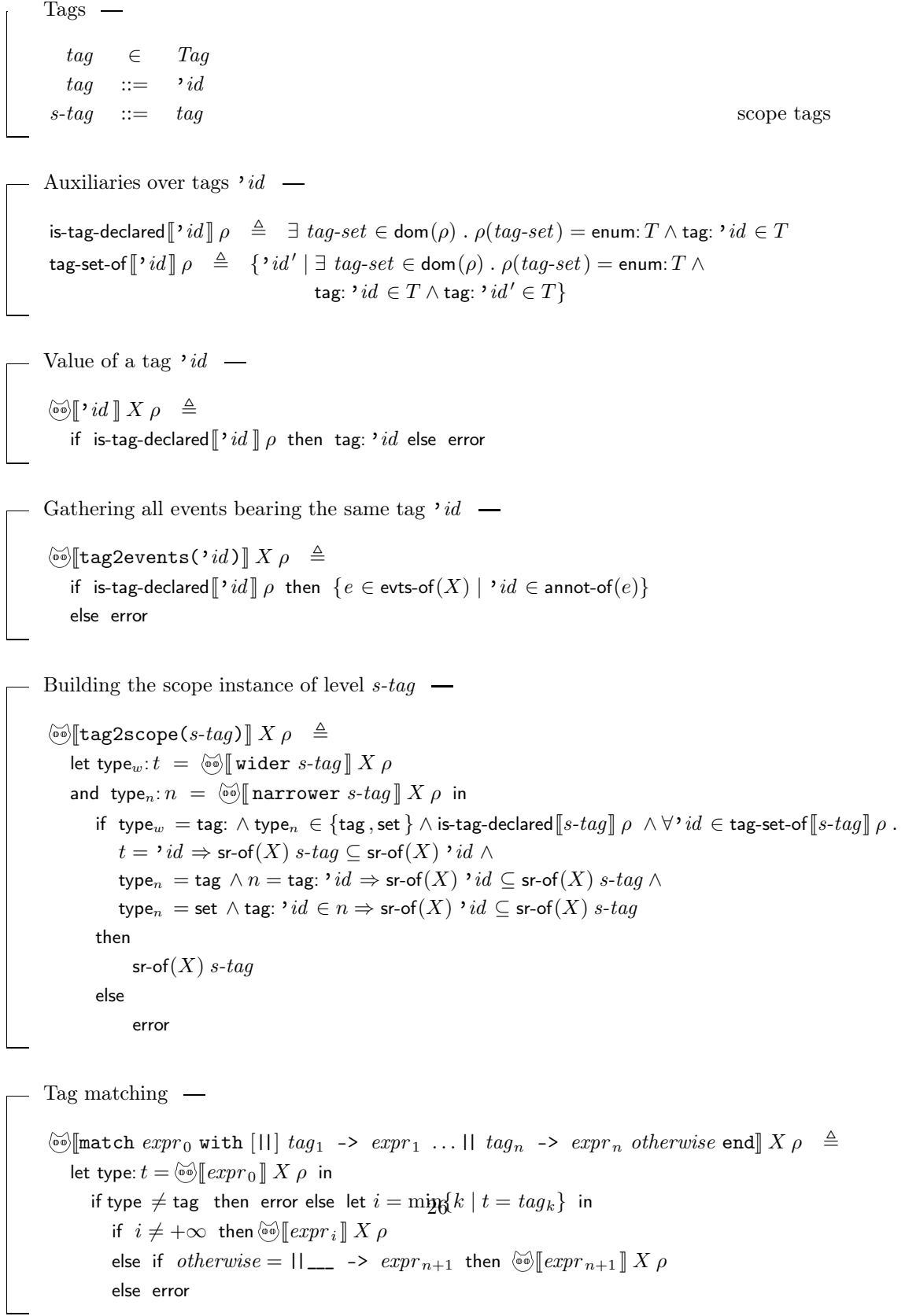
 
\begin{pcframed}[Tags] 
\begin{eqntabular}[fl]{@{\quad}rcl}
\mtag  &\in&\setofnonterminal{tag}\nonumber\\
%
%
\defmtag  &\grdef& \texttt{'}  \nonterminal{id}\nonumber\\
\defstag &\grdef& \mtag  \renumber{scope tags}
\end{eqntabular} 
\end{pcframed}

\begin{pcframed}[Auxiliaries over tags  \texttt{'}\nonterminal{id}]
\begin{eqntabular*}[fl]{@{\quad}l}
\defcatistagdeclared[\texttt{'}\nonterminal{id}]\:\catenvironment
\colsep{\triangleq}
\exists\
\nonterminal{tag-set}\in\domainof{\catenvironment}\suchthat\catenvironment(\nonterminal{tag-set})={\cattype{enum}}
T\wedge \cattypetag\texttt{'}\nonterminal{id} \in T
\\
\defcattagsetof[\texttt{'}\nonterminal{id}]\:\catenvironment
\colsep{\triangleq}
\{\texttt{'}\nonterminal{id}'\mid\begin{array}[t]{@{}l@{}}
\exists\
\nonterminal{tag-set}\in\domainof{\catenvironment}\suchthat\catenvironment(\nonterminal{tag-set})={\cattype{enum}}
T\wedge{}\\[0.33ex]
\cattypetag\texttt{'}\nonterminal{id} \in T
\wedge \cattypetag\texttt{'}\nonterminal{id}' \in T\}
\end{array}
\end{eqntabular*}
\end{pcframed}

\begin{pcframed}[Value of a tag \texttt{'}\nonterminal{id}]
\begin{eqntabular}[fl]{@{\quad}l}
\catsemantics[\texttt{'}
\nonterminal{id}\mskip2mu]\:\catconfiguration\:\catenvironment
\colsep{\triangleq}
\nonumber\\
\quad\iif\ \catistagdeclared[\texttt{'}
\nonterminal{id}\mskip2mu]\:\catenvironment\ \ithen\ \cattypetag\texttt{'}\nonterminal{id}\ \ielse\ \caterror\nonumber
\end{eqntabular}
\end{pcframed}


\begin{pcframed}[Gathering all events bearing the same tag $\texttt{'}\nonterminal{id}$]
\begin{eqntabular}[fl]{@{\quad}l}
\catsemantics[\defcattagtoevents\texttt{(}\texttt{'}\nonterminal{id} \texttt{)}]\:\catconfiguration\:\catenvironment
\colsep{\triangleq}\nonumber\\
\quad\iif\ \catistagdeclared[\texttt{'}\nonterminal{id}]\:\catenvironment\ \ithen\ \{\event\in\eventsof(\catconfiguration)\mid\texttt{'}\nonterminal{id} \in\tagsofevent(e)\}\nonumber\\
\quad\ielse\ \caterror\nonumber
\end{eqntabular}
\end{pcframed}

\begin{pcframed}[Building the scope instance of level $\stag$]
\begin{eqntabular}[fl]{@{\quad}l}
\catsemantics[\defcattagtoscope\texttt{(}\stag\texttt{)}]\:\catconfiguration\:\catenvironment
\colsep{\triangleq}\nonumber\\
\quad\llet\ \cattypenocolon{type}{$_w${:}}{}t\ =\ \catsemantics[\mskip3mu\defcatwider\ \stag\mskip2mu]\:\catconfiguration\:\catenvironment\nonumber\\
\quad\land\ \cattypenocolon{type}{$_n${:}}{}n\ =\ \catsemantics[\mskip3mu\defcatnarrower\ \stag\mskip2mu]\:\catconfiguration\:\catenvironment\ \lin\nonumber\\
%
\quad\quad\:\iif\ \begin{array}[t]{@{}l@{}}
\cattypenocolon{type}{$_w$}{}=\cattypetag
\wedge
\cattypenocolon{type}{$_n$}{}\in\{\cattypenocolon{tag}{},\cattypenocolon{set}{}\}
\wedge
\catistagdeclared[\stag]\:\catenvironment\ \wedge \forall \texttt{'}\nonterminal{id}\in\cattagsetof[\stag]\:\catenvironment\suchthat\\[0.33ex]
 t=\texttt{'}\nonterminal{id}
 \implies
 \srof(\catconfiguration)\:\stag\subseteq\srof(\catconfiguration)\:\texttt{'}\nonterminal{id}\wedge{}\\[0.33ex]
 \cattypenocolon{type}{$_n$}{}=\cattypenocolon{tag}{}\wedge
 n=\cattypetag\texttt{'}\nonterminal{id}
 \implies
 \srof(\catconfiguration)\:\texttt{'}\nonterminal{id}\subseteq\srof(\catconfiguration)\:\stag\wedge{}\\[0.33ex]
  \cattypenocolon{type}{$_n$}{}=\cattypenocolon{set}{}\wedge
 \cattypetag\texttt{'}\nonterminal{id}\in n
 \implies
 \srof(\catconfiguration)\:\texttt{'}\nonterminal{id}\subseteq\srof(\catconfiguration)\:\stag
 \end{array}
 \nonumber\\
\quad\quad\ithen\nonumber\\
\quad\quad\quad\quad\srof(\catconfiguration)\:\stag \nonumber\\
\quad\quad\:\ielse\nonumber\\
\quad\quad\quad\quad\caterror\nonumber 
\end{eqntabular} 
\end{pcframed} 
\begin{pcframed}[Tag matching]
\begin{eqntabular}[fl]{@{\quad}l}
\catsemantics\llbracket\defcatmatchtt\  \catexpr{}_0\  \catwithtt\  {[ \catalternative ]}\  \mtag _1\  \mathbin{\texttt{->}}\  \catexpr{}_1\ {}\ldots\catalternative\  \mtag _n\  \mathbin{\texttt{->}}\  \catexpr{}_n\  \textit{otherwise}\  \texttt{end}
\rrbracket\:\catconfiguration\:\catenvironment
\colsep{\triangleq}\nonumber\\
\quad\llet\ \cattype{type}t =
\catsemantics[\catexpr{}_0\mskip2mu]\:\catconfiguration\:\catenvironment
\ \lin\nonumber\\
\quad\quad\iif \cattypenocolon{type}{}\neq\cattypenocolon{tag}{}\ \ithen\ \caterror\ \ielse\ \llet\ i =\min\{k\mid t=\mtag _k\}\ \lin\nonumber\\
\quad\quad\quad\iif\ i\neq +\infty\ \ithen\catsemantics[\catexpr{}_i\mskip2mu]\:\catconfiguration\:\catenvironment\nonumber\\
\quad\quad\quad\ielse\ \iif\ \textit{otherwise}={\catalternative \mdefault\  \mathbin{\texttt{->}}\  \catexpr{}_{n+1}}\ \ithen\ \catsemantics[\catexpr{}_{n+1}\mskip2mu]\:\catconfiguration\:\catenvironment\nonumber\\
\quad\quad\quad\ielse\ \caterror\nonumber
\end{eqntabular}
\end{pcframed}
\vspace*{-2mm}
\caption{Tags and their semantics\label{fig:tags}
\label{fig:tag-sem}\label{fig:scopes} \label{fig:tag-match-sem}} \end{figure}

\paragraph{Tags}\label{Tags-sem} essentially are identifiers preceded by a quote \texttt{'} (to distinguish them form identifiers in bindings);
and we gather them in sets, as shown in Figure~\ref{fig:tags}. We first
define 
an auxiliary over a tag $\mtag$:
\begin{itemize}
\item $\catistagdeclared$ checks that $\mtag $ has been defined in an
environment $\catenvironment$, i.e.  belongs to an enumeration
$\nonterminal{tag-set}$ in $\catenvironment$.
\end{itemize} 

Now, the value of a tag \texttt{'}\nonterminal{id} is the corresponding typed
value if the tag has been declared in the environment $\catenvironment$, or an
$\caterror$ if not. 

Finally, the primitive \cattagtoevents\ gathers all events bearing the tag
$\mtag$, provided that the tag $\mtag$ is declared in the environment
$\catenvironment$.

\medskip

\afterpage{\clearpage}
\paragraph{Declarations.\quad} One can declare enumerations of tags named by an
identifier with the construct \catenum{}. One can use these tags to
\emph{annotate} \Lisa instructions, using the eponymous \catinstructions
construct.

\begin{figure}[!ht]
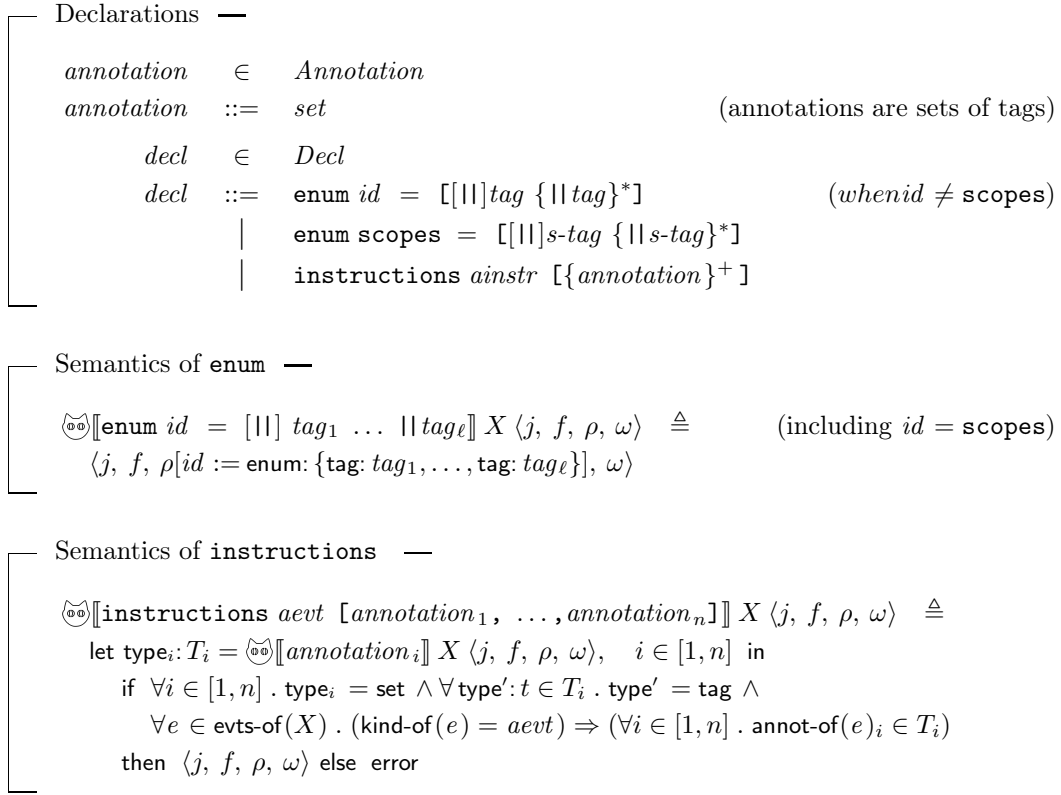

\vspace*{-2mm}
\begin{pcframed}[Declarations]
\begin{eqntabular}[fl]{@{\quad}rcl}
\nonterminal{annotation}&\in&\setofnonterminal{annotation}\nonumber\\
\defnonterminal{annotation} &\grdef& \catsetnonterminal\renumber{{(annotations are sets of tags)}}\\[1ex]
\nonterminal{decl}&\in&\setofnonterminal{decl}\nonumber\\
\defnonterminal{decl} 	&\grdef&	 \catenum{} \ \nonterminal{id}  
\  = \ \mathopen{\texttt{[}}  [\catalternative]  \mtag \  \{ \catalternative  \mtag  \}^\ast \mathclose{\texttt{]}}\renumber{$(when \nonterminal{id} \neq  \catscopes)$}\\
 	&\gror&	 \catenum{} \ \defcatscopes\  = \ \mathopen{\texttt{[}}  [\catalternative]  \stag\  \{ \catalternative  \stag \}^\ast \mathclose{\texttt{]}}\nonumber\\
 	&\gror&	 \catinstructions\  \nonterminal{ainstr}\ {\mathopen{\texttt{[}}\{\nonterminal{annotation}\}^+\mskip3mu\mathclose{\texttt{]}}}\nonumber
\end{eqntabular}
\end{pcframed}

\begin{pcframed}[Semantics of \catenum{}]
\begin{eqntabular}[fl]{@{\quad}l}
\catsemantics[{\defcatenum{} \ \nonterminal{id}\  = \
{[\catalternative]}\  \mtag _1\ \ldots\ \catalternative
\mtag _{\ell}}]\:\catconfiguration\:\catinstructionresultof{j}{f}{\catenvironment}
{\communicationrelationidentifiers}\colsep{\triangleq}\renumber{(including $\nonterminal{id}=\catscopes$)}\\
\quad\catinstructionresultof{j}{f}{\catassign{\catenvironment}{\nonterminal{id}}{\cattype{enum}\{\cattypetag{\mtag _1,\ldots,\cattypetag{\mtag_{\ell}}\}}}}{\communicationrelationidentifiers}\nonumber
\end{eqntabular}
\end{pcframed}
%
\begin{pcframed}[Semantics of \catinstructions]
\begin{eqntabular*}[fl]{@{\quad}l}
\catsemantics[\defcatinstructions\ \nonterminal{aevt}\
{\mathopen{\texttt{[}}\nonterminal{annotation}_1\texttt{,}\
\ldots\texttt{,}\nonterminal{annotation}_n\mathclose{\texttt{]}}}]\:\catconfiguration\:
\catinstructionresultof{j}{f}{\catenvironment}{\communicationrelationidentifiers}
\colsep{\triangleq}\nonumber\\
\quad\llet\ \cattypenocolon{type}{$_i${:}}T_i=\catsemantics[\nonterminal{annotation}_i]\:\catconfiguration\:
\catinstructionresultof{j}{f}{\catenvironment}{\communicationrelationidentifiers},\quad i\in\interval{1}{n}\ \lin\\
\quad\quad\:\iif\ \forall i\in\interval{1}{n}\suchthat
\cattypenocolon{type}{$_i$}=\cattypenocolon{set}{}
\wedge
\forall\mskip3mu\cattypenocolon{type}{$'${:}}t\in T_i\suchthat \cattypenocolon{type}{$'$}=\cattypenocolon{tag}{}
\wedge{}\nonumber\\
\quad\quad\phantom{\:\iif\ }\forall\event\in\eventsof(\catconfiguration)\suchthat(\catkindofevent(\event)=\nonterminal{aevt})\implies{}
(\forall i\in\interval{1}{n}\suchthat\tagsofevent(\event)_i\in T_i)\nonumber\\
\quad\quad\ithen\ \catinstructionresultof{j}{f}{\catenvironment}{\communicationrelationidentifiers}\ \ielse\ \caterror
\end{eqntabular*}
%
\end{pcframed}
\vspace*{-2mm}
\caption{Declarations \label{fig:decls}}
\vspace*{-5mm}
\end{figure}

Declarations (see Figure~\ref{fig:decls}) augment the environment. The effect
of an \catenum{} declaration is to extend the environment with the
corresponding set of tags. In other terms, the semantics of $\catenum{} \
\nonterminal{id}\  = \ {[\catalternative]}\ \mtag _1\ \ldots\ \catalternative
\mtag _n$ is to augment the environment $\catenvironment$ with the set of typed
tags $\mtag_1, \ldots \mtag_n$, under the name $\nonterminal{id}$.

The semantics of an \texttt{instruction} declaration is as follows: if there is
a tag not in the environment, we raise an error; and if there is an event whose
$i^{\text{th}}$ tag is not in the  $i^{\text{th}}$ tag set, we raise an error.

\subsubsection{Scopes.\quad}\label{sec:scopes-sem}
Semantically, we distinguish \emph{scope tags} \nonterminal{s-tag} from other
tags, as shown in Figure~\ref{fig:tags}. Thus for \catenum{} declarations,
the identifier \catscopes{} is reserved to declare scopes. If an
\texttt{enum\ \catscopes} declaration is provided then two functions
\catnarrower{} and \catwider{} must be declared on scope tags, to define
the set of all possible scope hierarchies. 
Finally, the primitive \cattagtoscope{} builds the relation between events
coming from instructions that belong to the same scope (\viz, the scope
instances of that scope) $\,$---$\,$relatively to a scope tree appearing in the
original program. We give its semantics in Figure~\ref{fig:scopes}.

\paragraph{Matching over tags} is as follows:
\bgroup\ttfamily
\begin{eqntabular*}[fl]{@{\quad}L}
match \catexpr{} with\\
\ \ || \ \nonterminal{tag}$_1$\ ->\ \catexpr{}$_1$\\
\ \ || \ ...\\
\ \ || \ \nonterminal{tag}$_n$\ ->\ \catexpr{}$_n$\\
\ \ || \mdefault \ ->\ \catexpr{}$_d$\\
end
\end{eqntabular*}%
\egroup

The value of the \catmatchtt{} expression is computed as follow: first
evaluate \catexpr{} to some value \nonterminal{v}, which must be a tag
\texttt{t}.  Then \nonterminal{v} is compared with the tags
\nonterminal{tag}$_{1}$, $\dots$, \nonterminal{tag}$_{n}$, in that order. If
some tag pattern \nonterminal{tag}$_{i}$ equals \texttt{t}, then the value of
the \catmatchtt{} is the value of the corresponding expression
\nonterminal{expr$_{i}$}.  Otherwise, the value of the \catmatchtt{} is the
value of the default expression \catexpr{}$_d$.  As the default clause
\texttt{\mdefault -> \catexpr{}$_d$} is optional, the \catmatchtt{}
construct may fail in $\caterror$.  We give the semantics of matching over tags
in Figure~\ref{fig:tag-match-sem}.

\subsubsection{Functions}\label{sec:functions-sem}

\paragraph{Functions} are first class values, as reflected by the anonymous
function construct \catfuntt{} \catpat{} \texttt{->} \catexpr{}. We
call the expression \catexpr{} the \emph{body} of the function. A
function takes one argument \catpat{} only. When this argument is a
tuple, it may be destructured by a tuple pattern \texttt{(}\nonterminal{id}$_1$,
$\dots$, \nonterminal{id}$_n${\tt)}. 

\begin{figure}[!ht]
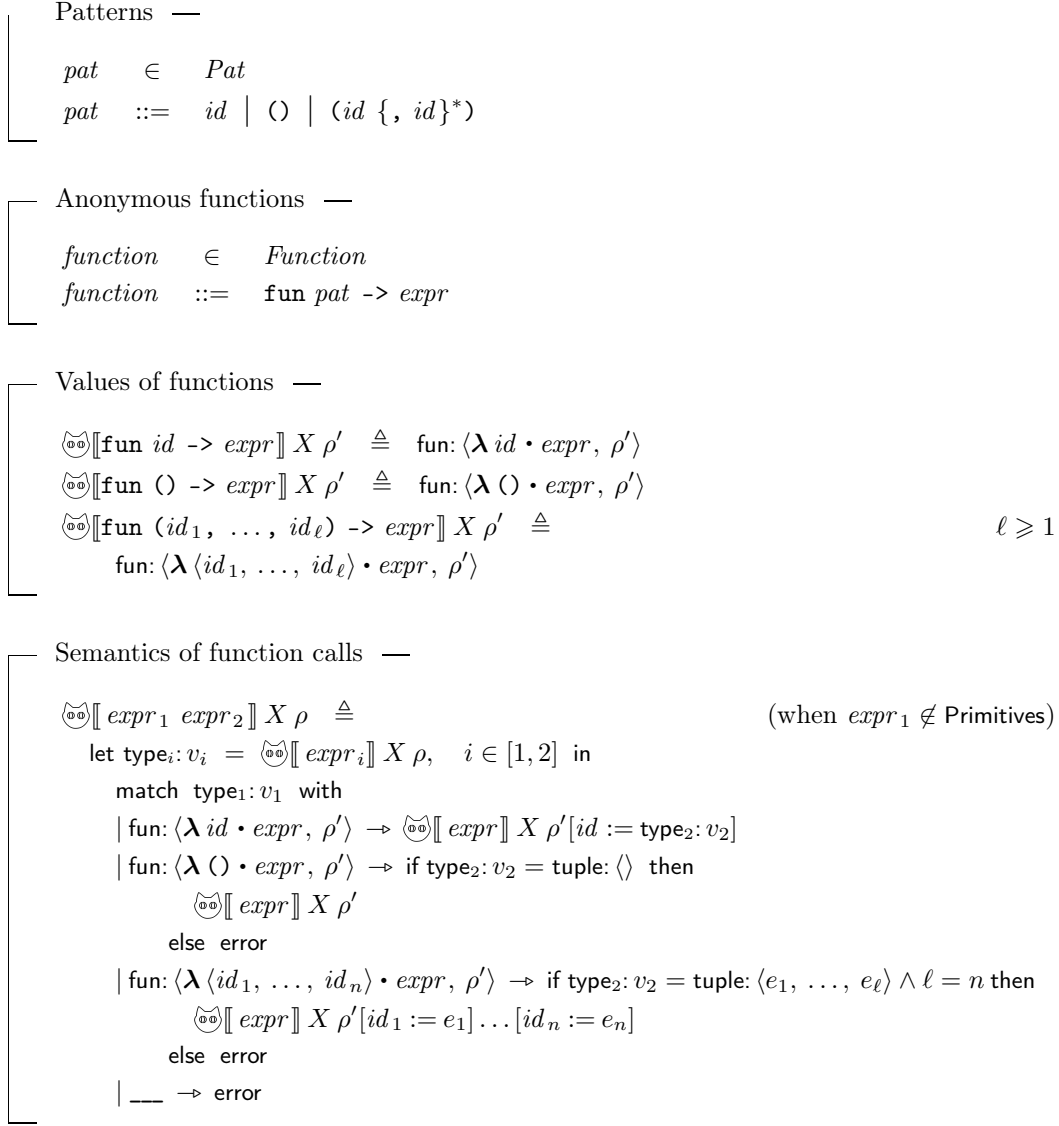

\begin{pcframed}[Patterns]
\begin{eqntabular}[fl]{@{\quad}rcl}
\catpat{}&\in&\setofnonterminal{pat}\nonumber\\
\defcatpat{} 	&\grdef&	 \nonterminal{id} \gror  \mathopen{\texttt{(}} \mathclose{\texttt{)}} \gror  \mathopen{\texttt{(}}  \nonterminal{id}\  \{ \texttt{,}\  \nonterminal{id} \}^\ast  \mathclose{\texttt{)}} \nonumber
\end{eqntabular}
\end{pcframed}
\begin{pcframed}[Anonymous functions]
\begin{eqntabular*}[fl]{@{\quad}rcl} 
\defcatfunction &\in&\catsetofallfunctions\\ 
\catfunction &\grdef& \catfuntt\  \catpat{}\  \texttt{->}\  \catexpr{}  \nonumber
\end{eqntabular*} 
\end{pcframed}

\begin{pcframed}[Values of functions]
\begin{eqntabular}[fl]{@{\quad}l}
\catsemantics[\defcatfuntt\  \nonterminal{id}\  \texttt{->}\  \catexpr{}]\:\catconfiguration\:\catenvironment'
\colsep{\triangleq}
\cattypefun\pair{\LAMBDA{\nonterminal{id}}\catexpr{}}{\catenvironment'}\nonumber\\
\catsemantics[\catfuntt\ \mathopen{\texttt{(}}\mathclose{\texttt{)}}\  \texttt{->}\  \catexpr{}]\:\catconfiguration\:\catenvironment'
\colsep{\triangleq}
\cattypefun\pair{\LAMBDA{\mathopen{\texttt{(}}\mathclose{\texttt{)}}}\catexpr{}}{\catenvironment'}\nonumber\\
\catsemantics[\catfuntt\  \mathopen{\texttt{(}}  \nonterminal{id}_1\texttt{,}\ \ldots\texttt{,}\  \nonterminal{id}_{\ell}  \mathclose{\texttt{)}}\  \texttt{->}\  \catexpr{}]\:\catconfiguration\:\catenvironment'
\colsep{\triangleq}\renumber{$\ell\geqslant1$}\\
\quad\quad\cattypefun\pair{\LAMBDA{\triple{\nonterminal{id}_1}{\ldots}{\nonterminal{id}_{\ell}}}\catexpr{}}{\catenvironment'}\nonumber
\end{eqntabular}
\end{pcframed}
\begin{pcframed}[Semantics of function calls]
\begin{eqntabular}[fl]{@{\quad}l}
\catsemantics[\mskip3mu\catexpr{}_1\  \catexpr{}_2\mskip2mu]\:\catconfiguration\:\catenvironment
\colsep{\triangleq}\renumber{(when $\catexpr{}_1\not\in\sffont{Primitives}$)}\\
\quad\llet\ \cattypenocolon{type}{$_i${:}}v_i\ =\ \catsemantics[\mskip3mu\catexpr{}_i]\:\catconfiguration\:\catenvironment,\quad i\in\interval{1}{2}\ \lin\nonumber\\
\quad\quad\mmatch\ \cattypenocolon{type}{$_1${:}}v_1\ \mwith\nonumber\\
\quad\quad\mcase \cattypefun\pair{\LAMBDA{\nonterminal{id}}\catexpr{}}{\catenvironment'}\mthen 
\catsemantics[\mskip3mu\catexpr{}]\:\catconfiguration\:\catassign{\catenvironment'}{\nonterminal{id}}{\cattypenocolon{type}{$_2${:}}v_2}\nonumber\\
\quad\quad\mcase \cattypefun\pair{\LAMBDA{\mathopen{\texttt{(}}\mathclose{\texttt{)}}}\catexpr{}}{\catenvironment'}\mthen \iif 
\cattypenocolon{type}{$_2${:}}v_2=\cattypetuple\singleton{}\ \ithen\nonumber\\
\quad\quad\quad\quad\quad\catsemantics[\mskip3mu\catexpr{}]\:\catconfiguration\:\catenvironment'\nonumber\\
\quad\quad\quad\quad\ielse\ \caterror\nonumber\\
\quad\quad\mcase \cattypefun\pair{\LAMBDA{\triple{\nonterminal{id}_1}{\ldots}{\nonterminal{id}_{n}}}\catexpr{}}{\catenvironment'}\mthen\iif\cattypenocolon{type}{$_2${:}}v_2=\cattypetuple{\triple{e_1}{\ldots}{e_{\ell}}}\wedge \ell= n\ithen\nonumber\\
\quad\quad\quad\quad\quad\catsemantics[\mskip3mu\catexpr{}]\:\catconfiguration\:
\catassign{\catassign{\catenvironment'}{\nonterminal{id}_1}{e_1}\ldots}{\nonterminal{id}_{n}}{e_{n}}\nonumber\\
\quad\quad\quad\quad\ielse\ \caterror\nonumber\\
\quad\quad\mcase \mdefault\mthen\caterror\nonumber
\end{eqntabular}
\end{pcframed}
\caption{Semantics of functions\label{fig:fun-sem}}
\end{figure}

The value of a function, given in Figure~\ref{fig:fun-sem}, is its
\emph{closure}; we write
$\pair{\LAMBDA{\nonterminal{id}}\nonterminal{body}}{\catenvironment'}$ for a
closure with parameter \nonterminal{id}, body \nonterminal{body} and
declaration environment $\catenvironment'$ (this closure can also be understood
as a triple $\triple{\nonterminal{id}}{\nonterminal{body}}{\catenvironment'}$).


\paragraph{Function calls} are written \catexpr{}$_1$
\catexpr{}$_2$. That is, functions are of arity one and the application
operator is left implicit. Notice that function application binds tighter than
all binary operators (see Section~\ref{sec:bin-ops}) and looser that postfix
operators (see Section~\ref{sec:post-ops}).  Furthermore the implicit
application operator is left-associative.

The \cat{} language has call-by-value semantics. That is, the effective
parameter \catexpr{}$_2$ is evaluated before being bound to the
function formal parameter, see Figure \ref{fig:fun-sem}. 

\subsubsection{Sets and tuples.}\label{sec:sets-tuples-sem}

\paragraph{Sets} are written as follows: \texttt{\{} \catexpr{}$_1$,
\catexpr{}$_2$, $\dots$, \catexpr{}$_n$\texttt{\}} with $n$
greater than $0$.  As events are not values, one cannot build a set of events
using explicit set expressions. Sets are homogeneous, \ie contain elements of
the same type. We give their semantics in Figure~\ref{fig:sets-and-tuples-sem}.
The value of \texttt{\{\}} is the empty set, and the value of
$\mathopen{\texttt{\{}}\catexpr{}_1\texttt{,} \ldots\texttt{,}\
\catexpr{}_n \mathclose{\texttt{\}}}$ is the set of values $\{v_1
\ldots v_n\}$ where the $v_i$ are the values of $\catexpr{}_i$.

\paragraph{Matching over sets}\label{sec:set-match} is as follows:
\bgroup\ttfamily
\begin{eqntabular*}[fl]{@{\quad}L}
match \catexpr{} with\\
\ \ || \ \{\} \ ->\ \catexpr{}$_1$\\
\ \ || \ \nonterminal{id}$_1$\ ++\ \nonterminal{id}$_2$\ ->\ \catexpr{}$_2$\\
end
\end{eqntabular*}%
\egroup

We compute the value of the \catmatchtt{} as follow: first evaluate
\catexpr{} to some value \nonterminal{v}, which must be a set. If
\nonterminal{v} is the empty set \texttt{\{\}}, then the value of the
\catmatchtt{} is the value of \catexpr{}$_1$.  Otherwise, if
\nonterminal{v} is a non-empty set $S$, then let $e$ be some element in $S$ and
$S'$ be the set $S$ minus the element $e$. The value of the \catmatchtt{} is
the value of \catexpr{}$_2$ in a context where \nonterminal{id}$_1$ is
bound to $e$ and \nonterminal{id}$_2$ is bound to $S'$. We give the semantics
of matching over sets in Figure~\ref{fig:sets-and-tuples-sem}, where the
non-deterministic choice $e\in s$ is arbitrary (and unknown). So the semantics in Figure~\ref{fig:sets-and-tuples-sem} returns one possible match (as opposed to all possibilities).

\paragraph{Tuples} include the empty tuple \texttt{()}, and constructed tuples
\texttt{(}\catexpr{}$_1$, \catexpr{}$_2$, $\dots$,
\catexpr{}$_n$\texttt{)}, with $n$ greater than $2$. In other words there
is no tuple of size one (which avoids ambiguity with grouping between parentheses).   

We give their semantics in Figure~\ref{fig:sets-and-tuples-sem}. The value of
\texttt{()} is the empty tuple $\singleton{}$, and the value of
$\triple{\catexpr{}_1}{\ldots}{\catexpr{}_n}$ is the tuple of
values $\triple{v_1}{\ldots}{v_n}$ where the $v_i$ are the values of
$\catexpr{}_i$; we do not impose that these values $\triple{v_1}{\ldots}{v_n}$ have the same type.

\paragraph{Grouping} is straightforward, as shown in
Figure~\ref{fig:sets-and-tuples-sem}: the semantics of a parenthesised
expression \texttt{(}\catexpr{}{\tt)} is the semantics of
\catexpr{}, idem for \texttt{begin} \catexpr{}\/ \texttt{end}.

\begin{figure}[!ht]
\begin{pcframed}[Sets and tuples]
\begin{eqntabular}[fl]{@{\quad}rcl}
\defcatsetnonterminal&\in&\setofnonterminal{Set}\nonumber\\
\catsetnonterminal&\grdef& \mathopen{\texttt{\{}} \mathclose{\texttt{\}}} \gror  \mathopen{\texttt{\{}}\catexpr{}\  \{\texttt{,}\  \catexpr{} \}^\ast\mathclose{\texttt{\}}}  \renumber{sets}\\[1ex]
\defcattuplenonterminal&\in&\setofnonterminal{Tuple}\nonumber\\
\cattuplenonterminal&\grdef& \mathopen{\texttt{(}} \mathclose{\texttt{)}} \gror  \mathopen{\texttt{(}}\catexpr{}\texttt{,}\  \catexpr{}\  \{\texttt{,}\  \catexpr{}\}^\ast\mathclose{\texttt{)}}  \renumber{tuples}
\end{eqntabular}
\end{pcframed}

\begin{pcframed}[Semantics of sets]
\begin{eqntabular}[fl]{@{\quad}l}
\catsemantics[\mathopen{\texttt{\{}}
\mathclose{\texttt{\}}}]\:\catconfiguration\:\catenvironment
\colsep{\triangleq}\cattypeset\emptyset\nonumber\\
\catsemantics[\mathopen{\texttt{\{}}\catexpr{}_1\texttt{,} \ldots\texttt{,}\  \catexpr{}_n \mathclose{\texttt{\}}}]\:\catconfiguration\:\catenvironment
\colsep{\triangleq}\renumber{$n\geqslant1$}\\
\quad\quad\llet\ \cattypenocolon{type}{$_i${:}}v_i\ =\ \catsemantics[\catexpr{}_i]\:\catconfiguration\:\catenvironment,\quad i\in\interval{0}{n}\ \lin\nonumber\\
%
%
\quad\quad\quad\llet\ S=\cattypeset\{\cattypenocolon{type}{$_1${:}}v_1,\ldots,\cattypenocolon{type}{$_n${:}}v_n\}\ \lin\nonumber\\
\quad\quad\quad\quad\iif\ \wellformedset{S}\ \ithen\ S\  \ielse\ \caterror\nonumber
\end{eqntabular}
\end{pcframed}

\begin{pcframed}[Set matching]
\begin{eqntabular}[fl]{@{\quad}l}
\catsemantics\llbracket\catmatchtt\  \catexpr{}_0\  \catwithtt\
{[ \catalternative ]}\  \mathopen{\texttt{\{}} \mathclose{\texttt{\}}}\
\mathbin{\texttt{->}}\  \catexpr{}_1 \catalternative  \
\nonterminal{id}_1  \mathbin{\texttt{++}}  \nonterminal{id}_2 \
\mathbin{\texttt{->}} \ \catexpr{}_2\
\texttt{end}\rrbracket\:\catconfiguration\:\catenvironment
\colsep{\triangleq}\nonumber\\
\quad\llet\ \cattype{type}s =
\catsemantics[\catexpr{}_0\mskip2mu]\:\catconfiguration\:\catenvironment
\ \lin\nonumber\\
\quad\quad\iif \cattypenocolon{type}{}\neq\cattypenocolon{set}{}\ \ithen\
\caterror\nonumber\\
\quad\quad\ielse\ \iif\ s=\emptyset\ \ithen\ \catsemantics[\catexpr{}_1\mskip2mu]\:\catconfiguration\:\catenvironment\nonumber\\
\quad\quad\ielse\ \llet\ e\in s\ \lin\nonumber\\
\quad\quad\quad\catsemantics[\catexpr{}_2\mskip2mu]\:\catconfiguration\:\catassign{\catassign{\catenvironment}{\nonterminal{id}_2}{\cattypeset (s\setminus\{e\})}}{\nonterminal{id}_1}{e}\nonumber
\end{eqntabular}
\end{pcframed}

\begin{pcframed}[Semantics of tuples] 
\begin{eqntabular}[fl]{@{\quad}l}
\catsemantics[\mathopen{\texttt{(}} \mathclose{\texttt{)}}]\:\catconfiguration\:\catenvironment
\colsep{\triangleq}\cattypetuple\singleton{}\nonumber\\
\catsemantics[\mathopen{\texttt{(}}\catexpr{}_1\texttt{,}\ \ldots\texttt{,}\ \catexpr{}_n\mathclose{\texttt{)}}]\:\catconfiguration\:\catenvironment
\colsep{\triangleq}\renumber{$n\geqslant2$}\\
\quad\quad\cattypetuple\triple{\catsemantics[\catexpr{}_1]\:\catconfiguration\:\catenvironment}{\ldots}{\catsemantics[\catexpr{}_n]\:\catconfiguration\:\catenvironment}\nonumber
\end{eqntabular}
\end{pcframed}

\begin{pcframed}[Semantics of grouping]
\begin{eqntabular}[fl]{@{\quad}l}
\catsemantics[\mathopen{\texttt{(}}\catexpr{}\mathclose{\texttt{)}}]\:\catconfiguration\:\catenvironment
\colsep{\triangleq}
\catsemantics[\catexpr{}]\:\catconfiguration\:\catenvironment\nonumber\\
\catsemantics[\mathopen{\texttt{begin}}\  \catexpr{}\  \mathclose{\texttt{end}}]\:\catconfiguration\:\catenvironment
\colsep{\triangleq}
\catsemantics[\catexpr{}]\:\catconfiguration\:\catenvironment\nonumber
\end{eqntabular}
\end{pcframed}
\caption{Semantics of sets, tuples and grouping\label{fig:sets-and-tuples-sem}} 
\end{figure}

\subsubsection{Bindings}\label{sec:bindings-sem} are of the form \catpat{} \texttt{=}
\catexpr{} or \nonterminal{id} \catpat{} \texttt{=}
\catexpr{}, where \nonterminal{id} \catpat{} \texttt{=}
\catexpr{} is syntactic sugar for \nonterminal{id} \texttt{=}
\catfuntt\ \catpat{} \texttt{->} \catexpr{}. As shown in
Figure~\ref{fig:bind}, bindings simply update the environment
$\catenvironment$. The bindings for \catpat{}{} \texttt{=}
\catexpr{} are as follows: if \catpat{}{} is \texttt{()}, then
\catexpr{} must evaluate to the empty tuple; if \catpat{}{} is
\nonterminal{id} or \texttt{(}\nonterminal{id}{\tt)}, then \nonterminal{id} is
bound to the value of \catexpr{}; if \catpat{} is a proper
tuple pattern \texttt{(}\nonterminal{id}$_1$, $\dots$, \nonterminal{id}$_n${\tt)}
with $n$ greater than $2$, then \catexpr{} must evaluate to a tuple
value of size $n$ \texttt{(} \nonterminal{v}$_1$, $\dots$,
\nonterminal{v}$_n$\texttt{)} and the names \nonterminal{id}$_1$, $\dots$,
\nonterminal{id}$_n$ are bound to the values \nonterminal{v}$_1$, $\dots$,
\nonterminal{v}$_n$. 

\begin{figure}[!ht]
\begin{pcframed}[Bindings]
\begin{eqntabular}[fl]{@{\quad}rcl}
\defcatbinding&\in&	\setofnonterminal{binding}\nonumber\\
\catbinding 	&\grdef& \catpat{} \mathbin{\texttt{=}}  \catexpr{} \gror  \nonterminal{id}\  \catpat{} \mathbin{\texttt{=}}   \catexpr{} \nonumber\\
& & \text{where}\ \nonterminal{id}\ \catpat{}\ \mathbin{\texttt{=}}\
\catexpr{} \colsep{\triangleq} \nonterminal{id}\ \mathbin{\texttt{=}}\
\catfuntt\  \catpat{}\  \texttt{->}\  \catexpr{} \nonumber
\end{eqntabular} \end{pcframed} 

\begin{pcframed}[Value of a binding]
\begin{eqntabular}[fl]{@{\quad}l}
\catsemantics[\mskip3mu\catpat{}\ \mathbin{\texttt{=}}\ \catexpr{}]\:\catconfiguration\:\catenvironment
\colsep{\triangleq}\mmatch \catpat{} \mwith\nonumber\\
\mcase \mathopen{\texttt{(}} \mathclose{\texttt{)}} \mthen \cattypetuple\singleton{}\nonumber\\
\mcase \nonterminal{id} \ \mcase \mathopen{\texttt{(}} \nonterminal{id} \mathclose{\texttt{)}} \mthen \catassign{\catenvironment}{\nonterminal{id}}{\catsemantics[\catexpr{}]\:\catconfiguration\:\catenvironment}\nonumber\\
\mcase \mathopen{\texttt{(}}  \nonterminal{id}_1\texttt{,}\ldots\texttt{,}\nonterminal{id}_m  \mathclose{\texttt{)}} \mthen \nonumber\\
\quad\mmatch(\catsemantics[\catexpr{}]\:\catconfiguration\:\catenvironment)\mwith\nonumber\\
\quad\quad\mcase \cattypetuple\triple{e_1}{\ldots}{e_m}\mthen\catassign{\catassign{\catenvironment}{\nonterminal{id}_1}{e_1}\ldots}{\nonterminal{id}_m}{e_m}\nonumber\\
\quad\quad\mcase \mdefault \mthen \caterror\nonumber\\[1ex]
\catsemantics[\mskip3mu\nonterminal{id}\ \catpat{}\ \mathbin{\texttt{=}}\
\catexpr{}]\:\catconfiguration\:\catenvironment
\colsep{\triangleq}
\catsemantics[\mskip3mu\nonterminal{id}\ \mathbin{\texttt{=}}\
\catfuntt\  \catpat{}\  \texttt{->}\  \catexpr{}]\:\catconfiguration\:\catenvironment
\nonumber
\end{eqntabular}
\end{pcframed}

\begin{pcframed}[Binding definitions]
\begin{eqntabular}[fl]{@{\quad}l}
\catsemantics[\texttt{let} \ \catbinding_1\ \texttt{and}\ \ldots \ \texttt{and}\  \catbinding_n]\:\catconfiguration\:\catenvironment
\colsep{\triangleq}
\renumber{$n\geqslant 1$}\\
\quad\quad(\catsemantics[\catbinding_n]\:\catconfiguration(\ldots(\catsemantics[\catbinding_1]\:\catconfiguration\:\catenvironment)\ldots))\nonumber
\end{eqntabular}

\vspace*{1mm}
\noindent$\ \:\,$Recursive function binding:
\begin{eqntabular}[fl]{@{\quad}l}
\catsemantics[\texttt{let}\ \texttt{rec}\ \nonterminal{id}_1\ \catpat{}_1\ \mathbin{\texttt{=}}\ \catexpr{}_1\ \texttt{and}\ \ldots \ \texttt{and}\ \nonterminal{id}_n\ \catpat{}_n\ \mathbin{\texttt{=}}\ \catexpr{}_n\mskip2mu]\:\catconfiguration\:\catenvironment
\colsep{\triangleq}\nonumber\\
\quad\llet\ \textit{cl}^\infty=\prod_{j=1}^{n}
\catsemantics[\catfuntt\  \catpat{}_i\  \texttt{->}\  \catexpr{}_i] 
\:\catconfiguration\:
{\catassign{\catassign{\catenvironment}{\nonterminal{id}_1}{\textit{cl}^\infty_1}\ldots}{\nonterminal{id}_n}{\textit{cl}^\infty_n}}\ \lin\nonumber\\
\quad\quad{\catassign{\catassign{\catenvironment}{\nonterminal{id}_1}{\textit{cl}^\infty_1}\ldots}{\nonterminal{id}_n}{\textit{cl}^\infty_n}}\nonumber
\end{eqntabular}

\vspace*{1mm}
\noindent $\ \:\,$Recursive set/relation binding:
\begin{eqntabular}[fl]{@{\quad}l}
\catsemantics[\texttt{let}\ \texttt{rec}\ \nonterminal{id}_1\ \mathbin{\texttt{=}}\ \catexpr{}_1\ \texttt{and}\ \ldots \ \texttt{and}\ \nonterminal{id}_n\ \mathbin{\texttt{=}}\ \catexpr{}_n\mskip2mu]\:\catconfiguration\:\catenvironment
\colsep{\triangleq}\nonumber\\
\quad\llet\ F_i(\prod_{j=1}^{n}x_j)\triangleq
\begin{array}[t]{@{}l}
\displaystyle\llet \ \cattypenocolon{type}{$_i${:}}s_i=\catsemantics[\catexpr{}_i]\:\catconfiguration\:\catassign{\catenvironment}{\prod_{j=1}^{n}\nonterminal{id}_j}{\prod_{j=1}^{n}x_j}\ \lin\\ 
\displaystyle\quad\iif (\cattypenocolon{type}{$_i$}\in\{\cattypenocolon{set}{},\cattypenocolon{rel}{}\})\ \ithen\  \cattypenocolon{type}{$_i${:}}s_i\ \ielse\ \caterror,\quad i\in\interval{1}{n}\\
\end{array}\nonumber\\
\quad\quad\lin\ \llet\ \prod_{i=1}^{n}e_i=\Lfp{\psubseteq}\LAMBDA{\prod_{i=1}^{n}x_i}\prod_{i=1}^{n}F_i(\prod_{i=1}^{n}x_i)\ \lin\nonumber\\
\quad\quad\quad\iif\ (\exists i\in\interval{1}{n}\suchthat e_i=\caterror)\ \ithen\ \caterror\ \ielse\ \catassign{\catassign{\catenvironment}{\nonterminal{id}_1}{e_1}\ldots}{\nonterminal{id}_n}{e_n}\nonumber
\end{eqntabular}
\end{pcframed}

\begin{pcframed}[Binding expressions]
\begin{eqntabular}[fl]{@{\quad}l}
\catsemantics[\texttt{let}\ \catbinding_1\ \texttt{and}\ \ldots \ \texttt{and}\ \catbinding_n\ \texttt{in}\  \catexpr{}]\:\catconfiguration\:\catenvironment
\colsep{\triangleq}\renumber{$n>1$}\\
\quad\catsemantics[\catexpr{}]\:\catconfiguration\:(\catsemantics[\texttt{let}\ \catbinding_1\ \texttt{and}\ \ldots \ \texttt{and}\ \catbinding_n]\:\catconfiguration\:\catenvironment)\nonumber
\end{eqntabular}

\begin{eqntabular}[fl]{@{\quad}l}
\catsemantics[\texttt{let}\ \texttt{rec}\ \nonterminal{id}_1\ \mathbin{\texttt{=}}\ \catexpr{}_1\ \texttt{and}\ \ldots \ \texttt{and}\ \nonterminal{id}_n\ \mathbin{\texttt{=}}\ \catexpr{}_n\ \texttt{in}\  \catexpr{}]\:\catconfiguration\:\catenvironment
\colsep{\triangleq}\nonumber\\
\quad\catsemantics[\catexpr{}]\:\catconfiguration\:(\catsemantics[\texttt{let}\ \texttt{rec}\ \nonterminal{id}_1\ \mathbin{\texttt{=}}\ \catexpr{}_1\ \texttt{and}\ \ldots \ \texttt{and}\ \nonterminal{id}_n\ \mathbin{\texttt{=}}\ \catexpr{}_n]\:\catconfiguration\:\catenvironment)\nonumber
\end{eqntabular}
\end{pcframed}
\vspace*{-2mm}
\caption{Bindings and their semantics \label{fig:bind}}
\end{figure}

%

\clearpage
\paragraph{Binding definitions}\label{sec:Binding-definitions-sem} happen through the \texttt{let} and \texttt{let
rec} constructs, which bind value names for the rest of a specification
evaluation. We give the semantics of binding definitions in
Figure~\ref{fig:bind}.

First, the construct \texttt{let} \catbinding$_1$ \texttt{and}
$\dots$ \texttt{and} \catbinding$_n$, that is, \texttt{let}
\catpat{}$_1$ \texttt{=} \catexpr{}$_1$ \texttt{and} $\dots$
\texttt{and} \catpat{}$_n$ \texttt{=} \catexpr{}$_n$, evaluates
\catexpr{}$_1$, $\dots$, \catexpr{}$_n$, and binds the names in
the patterns \catpat{}$_1$, $\dots$, \catpat{}$_n$ to the
resulting values.  

Second, for recursive function bindings \texttt{let rec} \nonterminal{id}$_1$
\catpat{}$_1$ \texttt{=} \catexpr{}$_1$ \texttt{and} $\dots$
\texttt{and} \nonterminal{id}$_e$  \catpat{}$_n$ \texttt{=}
\catexpr{}$_n$, we follow \cite{DBLP:journals/tcs/MilnerT91} where the
proof of existence and unicity of the infinite closure $\textit{cl}^\infty$ is
based on \cite{Aczel-88-non-wellfounded-sets}.

Third, for recursive set or relation bindings \texttt{let rec}
\nonterminal{id}$_1$ \texttt{=} \catexpr{}$_1$ \texttt{and} $\dots$
\texttt{and} \nonterminal{id}$_n$ \texttt{=} \catexpr{}$_n$, we compute
the least solution of the equations \nonterminal{id}$_1$ \texttt{=}
\catexpr{}$_1$, $\dots$, \nonterminal{id}$_n$ \texttt{=}
\catexpr{}$_n$ on sets or relations using inclusion for ordering. These fixpoint equations must satisfy the $\subseteq$-monotony (increasingness) hypotheses of \cite{Tarski55-fp} fixpoint theorem or else the result is undefined.

The recursive bindings may be mutually recursive. We suppose these recursive
definitions well-formed, \ie terminating.  The result of ill-formed definitions
is undefined (\ie an implementation might return an error or never terminate).

\paragraph{Binding expressions}\label{sec:bind-sem} happen through the
construct \texttt{let [rec]} \nonterminal{bindings} \texttt{in}
\catexpr{}, which locally binds the names defined by
\nonterminal{bindings} to evaluate \catexpr{}. Both non-recursive and
recursive bindings are allowed.

\subsubsection{Operators on sets and relations}\label{sec:operators-sets-relations-sem} Operators can be unary or
binary.  We list them in Figure~\ref{fig:ops}, and detail their semantics below. 
\begin{figure}[!ht]
\begin{pcframed}[Operators on sets and relations]
\begin{eqntabular}[fl]{@{\quad}rcl}
\defcatop&\in&\setofnonterminal{operators}\nonumber\\
\catop&\grdef&  \catexpr{} \mathbin{\texttt{++}} \catexpr{} \renumber{set addition}\\
        &\gror&  \catexpr{} \mathbin{\texttt{*}}  \catexpr{}  \renumber{cartesian product}\\
 	&\gror&	 \catexpr{} \mathbin{\texttt{|}} \catexpr{} \renumber{union}\\
        &\gror&  \catexpr{} \mathbin{\texttt{\&}} \catexpr{} \renumber{intersection}\\
        &\gror& \catexpr{} \mathbin{\texttt{;}} \catexpr{} \renumber{relation composition}\\
        &\gror&  \catexpr{} \mathbin{\texttt{+}} \renumber{transitive closure} \\
        &\gror&  \catexpr{}\ \mathord{\texttt{?}} \renumber{reflexive closure} \\
 	&\gror&	 \catexpr{} \mathbin{\texttt{*}} \renumber{reflexive and transitive closure} \\
        &\gror&  \catexpr{} \texttt{\^{}-1} \renumber{inverse}\\
	&\gror&  \catexpr{} \mathbin{\texttt{\textbackslash}} \catexpr{} \renumber{subtraction} \\
 	&\gror&	 \catnegation\,\catexpr{}  \renumber{complement}
\end{eqntabular}
\end{pcframed}
\caption{List of both unary and binary operators }\label{fig:ops}
\end{figure}

\medskip

\paragraph{Unary operators.\quad}\label{sec:post-ops} Given an expression
denoting a relation, we can build its identity closure with the operator
\texttt{?}, its reflexive-transitive closure with the operator \texttt{*}, its
transitive closure with \texttt{+}, its complement with $\catnegation$ and its
inverse with \texttt{\^{}-1}. These operators are postfix, and are defined on
relations only, except for the complement, which can apply to sets of events or
tags as well.  Figure~\ref{fig:unary-ops} gathers them all. We recall that the
value of identifier \catemptyrelation{} is the empty relation. 

\begin{figure}[!ht]
\begin{center}
\begin{eqntabular*}{@{\quad}c|c|L@{\quad}}
\textrm{postfixed\ }\texttt{op}&\defcatoperationof[\texttt{op}]\:\catconfiguration\: r& relation operation\\\cline{1-3}
\uLstrut\texttt{*}&\Lfp{\subseteq}\LAMBDA{x}\identityrelation_{\eventsof({\mbox{\footnotesize$\!\catconfiguration$}})}\mathbin{{\;}{\cup}{\;}} x\relationcomposition r&reflexive closure\\
\texttt{+}&\Lfp{\subseteq}\LAMBDA{x}r\mathbin{{\;}{\cup}{\;}} x\relationcomposition r&irreflexive closure\ulstrut\\
\texttt{?}&\identityrelation_{\eventsof({\mbox{\footnotesize$\!\catconfiguration$}})}{}\cup{}r&identity closure\ulstrut\\
\texttt{\^{}-1}&\{\pair{\event'}{\event}\mid \pair{\event}{\event'}\in r\}&inverse\ulstrut\\
\cline{1-3}
\end{eqntabular*}
\end{center}
\begin{pcframed}[Unary operators on relations]
\begin{eqntabular}[fl]{@{\quad}Ll}
%
&\catsemantics[\catexpr{}\
\texttt{op}]\:\catconfiguration\:\catenvironment
\colsep{\triangleq}\nonumber\\
&\quad\llet\ \cattype{type}r\ =\
\catsemantics[\catexpr{}]\:\catconfiguration\:\catenvironment\
\lin\nonumber\\
&\quad\quad\iif\cattypenocolon{type}{}=\cattypenocolon{rel}{}\ithen\nonumber\\
&\quad\quad\quad\cattyperel\catoperationof[\texttt{op}]\:\catconfiguration\:
r\nonumber\\
&\quad\quad\ielse\ \caterror\nonumber 
\end{eqntabular}
\end{pcframed}

\begin{pcframed}[Complement of a set or relation]
\begin{eqntabular}[fl]{@{\quad}l}
\quad\catsemantics[\catnegation\: \catexpr{}]\:\catconfiguration\:\catenvironment
\colsep{\triangleq}\nonumber\\
\quad\quad\llet\ \cattype{type}s\ =\ \catsemantics[\catexpr{}]\:\catconfiguration\:\catenvironment\ \lin\nonumber\\
\quad\quad\quad\iif\ \cattypenocolon{type}{}=\cattypenocolon{set}{}\ \wedge\ \forall\mskip3mu\cattype{type}e\in s\suchthat\cattypenocolon{type}{}=\cattypenocolon{evt}{}\ \ithen\nonumber\\
\quad\quad\quad\quad\cattypeset (\{\cattypeevt \event \mid \event \in \eventsof(\catconfiguration)\}\setminus s)\nonumber\\
\quad\quad\quad\ielse\iif
\ \cattypenocolon{type}{}=\cattypenocolon{set}{}\ \wedge\
\exists\mskip3mu\nonterminal{id}\in\domainof{\catenvironment}\suchthat\catenvironment(\nonterminal{id})=\cattype{enum} T\ \wedge\
 s\subseteq T\ \ithen\nonumber\\
\quad\quad\quad\quad\cattypeset (T\setminus s)\nonumber\\
%
%
%
%
\quad\quad\quad\ielse\ \iif\ \cattypenocolon{type}{}=\cattypenocolon{rel}{}\ \ithen\nonumber\\
\quad\quad\quad\quad\cattyperel ((\eventsof(\catconfiguration)\times\eventsof(\catconfiguration)) \setminus s)\nonumber\\
\quad\quad\quad\ielse\ \caterror\nonumber
\end{eqntabular}
\end{pcframed}

\caption{Semantics of unary operators~\label{fig:unary-ops}}
\end{figure}

\paragraph{Binary operators.\quad}\label{sec:bin-ops} 
We can build the sequence (or composition in the sense of
Figure~\ref{fig:aux-sem}) of two expressions with the operator \texttt{;},
defined on relations only.  We can add an element to a set: the addition
operator \catexpr{}$_1$ \texttt{++} \catexpr{}$_2$ operates on
sets. The value of \catexpr{}$_2$ must be a set of values $S$ and the
operator returns the set $S$ augmented with the value of
\catexpr{}$_1$.  We can build a new relation out of the cartesian
product of two sets of events, with the infix operator \texttt{*}. 

We can build the union, intersection, and difference of sets and relations as
summarised in Figure~\ref{fig:binary-sem}. The semantics of
$\catexpr{}_1 \mathbin{\texttt{op}} \catexpr{}_2$ is the
operator \texttt{op} applied to the sets (resp. relations) $s_1$ and $s_2$,
\viz, the values of $\catexpr{}_1$ and $\catexpr{}_2$.  

\begin{figure}[!ht]
\begin{pcframed}[Sequence of relations]
\begin{eqntabular}[fl]{@{\quad}l}
\catsemantics[\catexpr{}_1 \mathbin{\texttt{;}} \catexpr{}_2]\:\catconfiguration\:\catenvironment
\colsep{\triangleq}\nonumber\\
\quad\quad\llet\ \cattypenocolon{type}{$_i${:}}r_i\ =\ \catsemantics[\catexpr{}_i]\:\catconfiguration\:\catenvironment,\quad i\in\interval{1}{2}\ \lin\nonumber\\
\quad\quad\quad\iif\cattypenocolon{type}{$_1$}=\cattypenocolon{type}{$_2$}=\cattypenocolon{rel}{}\ithen\nonumber\\
\quad\quad\quad\quad\cattyperel r_1\relationcomposition r_2\nonumber\\
\quad\quad\quad\ielse\ \caterror\nonumber
\end{eqntabular}
\end{pcframed}

\begin{pcframed}[Adding an element to a set]
\begin{eqntabular}[fl]{@{\quad}l}
\catsemantics[\catexpr{}_1 \mathbin{\texttt{++}} \catexpr{}_2]\:\catconfiguration\:\catenvironment
\colsep{\triangleq}\nonumber\\
\quad\quad\llet\ \cattypenocolon{type}{$_i${:}}v_i\ =\ \catsemantics[\catexpr{}_i]\:\catconfiguration\:\catenvironment,\quad i\in\interval{1}{2}\ \lin\nonumber\\
%
%
\quad\quad\quad\llet\ S = \cattypeset (\{\cattypenocolon{type}{$_1${:}}v_1\}\cup v_2)\ \lin\nonumber\\
\quad\quad\quad\quad\iif\ \wellformedset{S}\ \ithen\ S\ \:\ielse\ \caterror\nonumber
\end{eqntabular}
\end{pcframed}

\begin{pcframed}[Cartesian product of two sets of events]
\begin{eqntabular}[fl]{@{\quad}l}
\catsemantics[\catexpr{}_1 \mathbin{\texttt{*}} \catexpr{}_2]\:\catconfiguration\:\catenvironment
\colsep{\triangleq}\nonumber\\
\quad\quad\llet\ \cattypenocolon{type}{$_i${:}}s_i\ =\ \catsemantics[\catexpr{}_i]\:\catconfiguration\:\catenvironment,\quad i\in\interval{1}{2}\ \lin\nonumber\\
%
\quad\quad\quad\iif\ \cattypenocolon{type}{$_1$}=\cattypenocolon{type}{$_2$}=\cattypenocolon{set}{}\wedge\forall\mskip3mu\cattype{type}v\in s_1\cup s_2\suchthat\cattypenocolon{type}{}=\cattypenocolon{evt}{}\
\ithen\nonumber\\
\quad\quad\quad\quad\cattyperel s_1\times s_2\nonumber\\
\quad\quad\quad\ielse\ \caterror\nonumber
\end{eqntabular}
\end{pcframed}

\begin{eqntabular*}{@{\quad}C|c|L@{\quad}}
\texttt{op}&\defcatbinaryoperationof[\texttt{op}]\\\cline{1-3}
\uLstrut\texttt{|}&\cup&union\\
\texttt{\&}&\cap&intersection\\
\texttt{\textbackslash}&\setminus&difference\\\cline{1-3}
\end{eqntabular*}

\begin{pcframed}[Binary operators relative to both sets and relations]
\begin{eqntabular}[fl]{@{\quad}l}
\catsemantics[\catexpr{}_1 \mathbin{\texttt{op}} \catexpr{}_2]\:\catconfiguration\:\catenvironment
\colsep{\triangleq}\nonumber\\
\quad\quad\llet\ \cattypenocolon{type}{$_i${:}}s_i\ =\ \catsemantics[\catexpr{}_i]\:\catconfiguration\:\catenvironment,\quad i\in\interval{1}{2}\ \lin\nonumber\\
\quad\quad\quad\iif\ \cattypenocolon{type}{$_1$}=\cattypenocolon{type}{$_2$}=\cattypenocolon{set}{}\wedge{}\wellformedset{s_1\cup s_2}\ \ithen\nonumber\\
%
\quad\quad\quad\quad\cattypeset s_1\mathbin{(\catbinaryoperationof[\texttt{op}])} s_2\nonumber\\
\quad\quad\quad\ielse\ \iif\cattypenocolon{type}{$_1$}=\cattypenocolon{type}{$_2$}=\cattypenocolon{rel}{}\ithen\nonumber\\
\quad\quad\quad\quad\cattyperel s_1\mathbin{(\catbinaryoperationof[\texttt{op}])} s_2\nonumber\\
\quad\quad\quad\ielse\ \caterror\nonumber
\end{eqntabular}
\end{pcframed}
\caption{Semantics of binary operators~\label{fig:binary-sem}}
\end{figure}

\clearpage
\subsection{Constraints}\label{sec:requirements} 
Constraints (see Figure~\ref{fig:reqs}) can be checks, procedure calls, or
iteration over sets.

\begin{figure}[!ht]
\begin{pcframed}[Constraints]
\begin{eqntabular*}[fl]{@{\quad}rcl} 
\defcatconstraint
 &\in&\catsetofallrequirements\\ 
\catconstraint
 &\grdef& \\ 
&\gror&	 \nonterminal{flagoption}\  \catcheck\  \catexpr{}\  \texttt{as}\
\nonterminal{id} \\
&\gror&	 \catcalltt
\ \nonterminal{id}\  \catexpr{} \\
%
%
%
&\gror&	 \catforalltt\ \nonterminal{id}\  \texttt{in}\  \catexpr{}\ \catdott\  \nonterminal{statements} \  \texttt{end}  
%
\\
\defnonterminal{flagoption}&\grdef&[\catflagtt]
\end{eqntabular*} 
\end{pcframed}

\caption{Constraints \label{fig:reqs}}
\end{figure}

\subsubsection{Checks}\label{sec:checks} happen through the construct
\catcheck\ \catexpr{}, which evaluates \catexpr{} and
applies the check \catcheck.  There are six checks: acyclicity
(keyword \catacyclictt), irreflexivity (keyword \catirreflexivett) and
emptyness (keyword \catemptytt); and their negations. If the check succeeds,
the candidate execution is allowed so far. Otherwise, the candidate execution
is forbidden.

A check can optionally be named \nonterminal{id}, using the keyword
\texttt{as}.  A check can also be flagged, by prefixing it with the
\catflagtt keyword.  Flagged checks must be named with the \texttt{as}
construct. Failed flagged checks do not stop evaluation; instead failed flagged
checks are recorded under their name in the component $f$ of the semantics of
the \cat{} specification, for example to handle flagged candidate
executions later within our \textsf{herd7} tool. We give the semantics of
checks in Figure~\ref{fig:checks-sem}. 

\begin{figure}[!ht]
\begin{pcframed}[Checks]
\begin{eqntabular*}[fl]{@{\quad}rcl} 
\defcatcheck 	&\grdef&	 \nonterminal{checkname} \gror
\catnegation\,\nonterminal{checkname}  \\
\defnonterminal{checkname} &\grdef&	 \catacyclictt \gror  \catirreflexivett \gror \catemptytt 
\end{eqntabular*} 
\end{pcframed}

\begin{eqntabular*}[fl]{@{\quad}l}
\defcatcheckconditiononrelation[\textit{check}]\ R
\colsep{\triangleq}\nonumber\\
\quad\mmatch\ \catcheck\ \mwith\nonumber\\
\quad\begin{array}[t]{@{}lcl@{}}
\mcase\phantom{\catnegation\,}\defcatacyclictt&\mthen&\catcheckconditiononrelation[\catirreflexivett]\ R^+\\
\mcase\phantom{\catnegation\,}\defcatirreflexivett&\mthen&(\forall \event\in\fieldof{R}\suchthat\pair{\event}{\event}\not\in R)\\
\mcase\phantom{\catnegation\,}\defcatemptytt&\mthen& R=\emptyset\\
\mcase\defcatnegation\,\textit{check}\mskip3mu'&\mthen&\ \neg(\catcheckconditiononrelation[\textit{check}\mskip3mu']\ R)
\end{array}\nonumber
\end{eqntabular*}

\begin{eqntabular*}[fl]{@{\quad}l}
\defcatcheckrelation[\mskip3mu\nonterminal{flagoption}\ \catcheck\ R\ \texttt{as}\  \nonterminal{id}]\ \catinstructionresultof{j}{f}{\catenvironment}{\communicationrelationidentifiers}
\colsep{\triangleq}\\
\quad\iif\ (\catcheckconditiononrelation[\textit{check}]\ R)\ \ithen\nonumber\\
\quad\quad\catinstructionresultof{j}{f}{\catenvironment}{\communicationrelationidentifiers}\nonumber\\
\quad\ielse\ \iif\ \nonterminal{flagoption}=\defcatflagtt\ \ithen\nonumber\\
\quad\quad\catinstructionresultof{j}{f\cup\{\nonterminal{id}\}}{\catenvironment}{\communicationrelationidentifiers}\nonumber\\
\quad\ielse\  \catinstructionresultof{\catforbidden}{f}{\catenvironment}{\communicationrelationidentifiers}\nonumber
\end{eqntabular*}
\begin{pcframed}[Value of checks]
\begin{eqntabular*}[fl]{@{\quad}l}
\catsemantics[\mskip3mu{\nonterminal{flagoption}\  \catcheck\  \catexpr{}\  \texttt{as}\  \nonterminal{id}}]\:\catconfiguration\:\catinstructionresultof{j}{f}{}{\catenvironment}
\colsep{\triangleq}\nonumber\\
\quad\llet\ \cattype{type}R = \catsemantics[\catexpr{}]\:\catconfiguration\:\catenvironment\ \lin\\
\quad\quad\iif\ (\cattypenocolon{type}{}=\cattypenocolon{rel}{})\vee(\cattypenocolon{type}{}=\cattypenocolon{set}{}\wedge\catcheck\in\{\catemptytt,\catnegation\,\catemptytt\})\ \ithen\ \nonumber\\
\quad\quad\quad\catcheckrelation[\mskip3mu\nonterminal{flagoption}\ \catcheck\ R\ \texttt{as}\  \nonterminal{id}]\ \catinstructionresultof{j}{f}{\catenvironment}{\communicationrelationidentifiers}\nonumber\\
\quad\quad\ielse\ \caterror\nonumber\\
\end{eqntabular*}
\end{pcframed}

\caption{Checks and their semantics\label{fig:checks-sem}}
\end{figure}

Flagged checks are useful for specifications with statements that impact the
semantics of an entire program, \eg, in the case of specifications phrased in
terms of data races, such as C++~\cite{DBLP:journals/corr/WickersonBD15} or
HSA~\cite{HSA-Foundation-PSAS-2015}.

\afterpage{\clearpage}
\subsubsection{Procedures}\label{sec:procedures-sem}

\paragraph{Procedures} \label{sec:procedures} have no result and cannot be recursive: the body of a
procedure is a list of statements and the procedure will be invoked to apply
the constraints within its body. Intended usage of procedures is to define
constraints that are checked later. Figure~\ref{fig:proc-sem} gives the
semantics of procedures: just like functions, procedure declarations simply
augment the environment $\catenvironment$ with their closure.

\paragraph{Procedure calls} are written $\catcalltt
\  \nonterminal{id}\
\catexpr{}$, where $\nonterminal{id}$ is the name of a previously
defined procedure. The bindings performed during the call of a procedure are
discarded when the procedure returns, all other effects (\eg checks or flags,
see Section~\ref{sec:checks}) performed are retained.  Procedures cannot be
recursive.
\begin{figure}[!ht]
\begin{pcframed}[Procedures]
\begin{eqntabular*}[fl]{@{\quad}rcl} 
\defcatprocedure &\in&\catsetofallprocedures\\ 
\catprocedure &\grdef& \catprocedurett
\  \nonterminal{id}\ pat\  = \ \{ \catinstruction \}^+\  \texttt{end}  \\
\end{eqntabular*} 
\end{pcframed}

\begin{pcframed}[Declarations of procedures]
\begin{eqntabular*}[fl]{@{\quad}l}
\catsemantics[\defcatprocedurett
\  \nonterminal{id}_1\  \nonterminal{id}_2\  = \ \{ \catinstruction \}^+\  \texttt{end}]\:\catconfiguration\:\catinstructionresultof{j}{f}{\catenvironment}{\communicationrelationidentifiers}
\colsep{\triangleq}\nonumber\\
\quad\quad\catinstructionresultof{j}{f}{\catassign{\catenvironment}{\nonterminal{id}_1}{\cattypeproc\pair{\LAMBDA{\nonterminal{id}_2}\{ \catinstruction \}^+}{\catenvironment}}}{\communicationrelationidentifiers}\nonumber\\
\catsemantics[\catprocedurett
\  \nonterminal{id}\ \mathopen{\texttt{(}}\mathclose{\texttt{)}}\ = \ \{ \catinstruction \}^+\  \texttt{end}]\:\catconfiguration\:\catinstructionresultof{j}{f}{\catenvironment}{\communicationrelationidentifiers}
\colsep{\triangleq}\nonumber\\
\quad\quad\catinstructionresultof{j}{f}{\catassign{\catenvironment}{\nonterminal{id}}{\cattypeproc\pair{\LAMBDA{\mathopen{\texttt{(}}\mathclose{\texttt{)}}}\{ \catinstruction \}^+}{\catenvironment}}}{\communicationrelationidentifiers}\nonumber\\
\catsemantics[\catprocedurett
\  \nonterminal{id}\ \mathopen{\texttt{(}}  \nonterminal{id}_1\texttt{,}\ \ldots\texttt{,}\  \nonterminal{id}_{\ell}  \mathclose{\texttt{)}}\ = \ \{ \catinstruction \}^+\  \texttt{end}]\:\catconfiguration\:\catinstructionresultof{j}{f}{\catenvironment}{\communicationrelationidentifiers}
\colsep{\triangleq}\nonumber\\
\quad\quad\catinstructionresultof{j}{f}{\catassign{\catenvironment}{\nonterminal{id}}{\cattypeproc\pair{\LAMBDA{\triple{\nonterminal{id}_1}{\ldots}{\nonterminal{id}_{\ell}}}\{ \catinstruction \}^+}{\catenvironment}}}{\communicationrelationidentifiers}\nonumber\\
\end{eqntabular*}
\end{pcframed}
\begin{pcframed}[Procedure calls]
\begin{eqntabular*}[fl]{@{\quad}l}
\catsemantics[{\defcatcalltt
\  \nonterminal{id}\  \catexpr{}}]\:\catconfiguration\:\catinstructionresultof{j}{f}{\catenvironment}{\communicationrelationidentifiers}
\colsep{\triangleq}\nonumber\\
\quad\iif\ \nonterminal{id}\not\in\domainof{\catenvironment}\ \ithen\ \caterror\ \ielse\nonumber\\
\quad\mmatch\ \catenvironment(\nonterminal{id})\ \mwith\nonumber\\
\quad\quad\mcase \pair{\LAMBDA{\nonterminal{id}_1}\{ \catinstruction \}^+}{\catenvironment'}\mthen\nonumber\\
\quad\quad\quad\quad\catsemantics[\{ \catinstruction \}^+]\:\catconfiguration\:\catassign{\catenvironment'}{\nonterminal{id}_1}{(\catsemantics[\catexpr{}]\:\catconfiguration\:{\catenvironment})}\nonumber\\
\quad\quad\mcase \pair{\LAMBDA{\mathopen{\texttt{(}}\mathclose{\texttt{)}}}\{ \catinstruction \}^+}{\catenvironment'}\mthen\nonumber\\
\quad\quad\quad\quad\llet\ \cattype{type}v\ =\ \catsemantics[\catexpr{}]\:\catconfiguration\:\catinstructionresultof{j}{f}{\catenvironment}{\communicationrelationidentifiers}\ \lin\nonumber\\
\quad\quad\quad\quad\quad\iif\cattype{type}v=\cattypetuple\singleton{}\ \ithen\\
\quad\quad\quad\quad\quad\quad\catsemantics[\{ \catinstruction \}^+]\:\catconfiguration\:\catinstructionresultof{j}{f}{\catenvironment'}{\communicationrelationidentifiers}\nonumber\\
\quad\quad\quad\quad\quad\ielse\ \caterror\nonumber\\
\quad\quad\mcase \pair{\LAMBDA{\triple{\nonterminal{id}_1}{\ldots}{\nonterminal{id}_n}}\{ \catinstruction \}^+}{\catenvironment'}\mthen\nonumber\\
\quad\quad\quad\quad\llet\ \cattype{type}v\ =\ \catsemantics[\catexpr{}]\:\catconfiguration\:\catinstructionresultof{j}{f}{\catenvironment}{\communicationrelationidentifiers}\ \lin\nonumber\\
\quad\quad\quad\quad\quad\iif\cattype{type}v=\cattypetuple{\triple{e_1}{\ldots}{e_{\ell}}}\wedge (n=\ell)\ithen\nonumber\\
\quad\quad\quad\quad\quad\quad\catsemantics[\{ \catinstruction \}^+]\:\catconfiguration\:\catinstructionresultof{j}{f}{
\catassign{\catassign{\catenvironment'}{\nonterminal{id}_1}{e_1}\ldots}{\nonterminal{id}_n}{e_n}}{\communicationrelationidentifiers}\nonumber\\
\quad\quad\quad\quad\quad\ielse\ \caterror\nonumber\\
\quad\quad\mcase \mdefault\mthen\caterror\nonumber\end{eqntabular*}
\end{pcframed}
\vspace*{-2mm}
\caption{Semantics of procedures\label{fig:proc-sem}}
\end{figure}

\subsubsection{Iteration over sets}\label{sec:iteration-over-sets} We can
iterate checks over sets with the \catforalltt construct: 

\bgroup\ttfamily
\begin{eqntabular*}[fl]{@{\quad}L}
forall \nonterminal{id} in \catexpr{} do\\
\ \ \nonterminal{statements}\\
end
\end{eqntabular*}%
\egroup

The expression \catexpr{} must evaluate to a set $S$. Then, the list of
statements \nonterminal{statements} is evaluated for all bindings of the name
\nonterminal{id} to some element $e$ of $S$.  In practice, as failed checks
forbid the candidate execution, this amounts to checking the conjunction of the
checks within \nonterminal{statements} for all the elements of $S$.  Similarly
to procedure calls, the bindings performed during an iteration are discarded
when iteration ends, all other cumulated effects (e.g. checks) being retained.  We give
the semantics of iteration in Figure~\ref{fig:iter-sem}; the iteration is
non-deterministic since the choice $e$ in $S$ is arbitrary and unknown.

\begin{figure}[!ht]
\begin{eqntabular*}[fl]{@{\quad}l}
\defcatiterate\ \nonterminal{id}\ S\ \{ \catinstruction \}^+ \ \catconfiguration\:\catinstructionresultof{j}{f}{\catenvironment}{\communicationrelationidentifiers}
\colsep{\triangleq}\nonumber\\
\quad\iif\ S=\emptyset\ \ithen \catinstructionresultof{j}{f}{\catenvironment}{\communicationrelationidentifiers}\nonumber\\
\quad\ielse\ \llet\ e\in S\ \lin\nonumber\\
\quad\quad\llet\ r = \catsemantics[\{  \catinstruction \}^+]\:\catconfiguration\:\catinstructionresultof{j}{f}{\catassign{\catenvironment}{\nonterminal{id}}{e}}{\communicationrelationidentifiers}\ \lin\nonumber\\
\quad\quad\quad\iif\ r = \caterror\ \ithen\ \caterror\ \ielse\nonumber\\
\quad\quad\quad\llet\ \catinstructionresultof{j'}{f'}{\catenvironment'}{\communicationrelationidentifiers}=\ r\ \lin\nonumber\\
\quad\quad\quad\quad\iif\ j'= \catallowed\ \ithen\\
\quad\quad\quad\quad\quad\catiterate\ \nonterminal{id}\ (S\setminus\{e\})\ \{ \catinstruction \}^+ \ \catconfiguration\:\catinstructionresultof{j'}{f'}{\catenvironment'}{\communicationrelationidentifiers}\nonumber\\
\quad\quad\quad\quad\ielse\ \catinstructionresultof{j'}{f'}{\catenvironment'}{\communicationrelationidentifiers}\nonumber
\end{eqntabular*}

\begin{pcframed}[Iteration over sets]
\begin{eqntabular*}[fl]{@{\quad}l}
\catsemantics[\defcatforalltt\  \nonterminal{id}\  \texttt{in}\  \catexpr{}\ \catdott\  \{ \catinstruction \}^+ \  \texttt{end}]\:\catconfiguration\:\catinstructionresultof{j}{f}{\catenvironment}{\communicationrelationidentifiers}
\colsep{\triangleq}\nonumber\\
\quad\llet\ \cattype{type}S\ =\ \catsemantics[\catexpr{}]\:\catconfiguration\:\catinstructionresultof{j}{f}{\catenvironment}{\communicationrelationidentifiers}\ \lin\nonumber\\
\quad\quad\iif\ \cattypenocolon{type}{}\neq\cattypenocolon{set}{}\ \ithen\ \caterror\nonumber\\
\quad\quad\ielse\ 
\catiterate\ \nonterminal{id}\ S\ \{ \catinstruction \}^+ \ \catconfiguration\:\catinstructionresultof{j}{f}{\catenvironment}{\communicationrelationidentifiers}\nonumber
\end{eqntabular*}
\end{pcframed}
\caption{Semantics of iteration~\label{fig:iter-sem}}
\end{figure}

\afterpage{\clearpage}
\subsection{Requirements}\label{sec:Requirements}

Requirements are the constitutive blocks of a \cat{} specification. Their
evaluation goes as given in Figure~\ref{fig:spec-requirements}.
Requirements can be statements, or \catwithtt bindings.

\subsubsection{Statements} and their semantics have been presented in the
sections above. At the level of requirements, we evaluate lists of statements,
gather their evaluation $\catinstructionresultof{j}{f}{\catenvironment}{\communicationrelationidentifiers}$, and the final verdict forgets the environment $\catenvironment$ to built the result (see Figure \ref{fig:spec-sem}).

\subsubsection{Candidate extension via \protect\catwithtt{}
binding}\label{sec:candidate-execution-extension} happens through the construct
\catwithtt \nonterminal{id} \catfromtt \catexpr{}. This construct
extends the current environment by one binding (see Figure
\ref{fig:bind-with}). The grammar only allows \catwithtt bindings to occur
at the top-level. The expression \catexpr{} is evaluated to a set $S$.
Then the remainder of the specification is evaluated for each choice of element
$e$ in $S$ in an environment extended by a binding of the name \nonterminal{id}
to $e$.

The final verdict at top level in Figure~\ref{fig:spec-sem} gets rid of the environment and returns the communication relations obtained by finding the value of the communication relation identifiers $\nonterminal{id}$ in the environment.


\begin{figure}[!ht]
\begin{pcframed}[Requirements]
\begin{eqntabular}[fl]{@{\quad}rcl}
\defcatrequirements&\in&\catrequirements\nonumber\\
\catrequirements &\grdef&\catinstruction \nonumber\\
&\gror&\catinstruction \ \catrequirements\nonumber\\
&\gror& \catwithtt\  \nonterminal{id}\  \catfromtt\  \catexpr{}\ \catrequirements\nonumber
\end{eqntabular}

\begin{eqntabular}[fl]{@{\quad}l}
\catsemantics[\catrequirements]\colsep{\in}\setofallcatconfigurations\rightarrow\catinstructionsemanticdomain\rightarrow\wp(\catinstructionsemanticdomain)
\nonumber\\[1ex]
\catsemantics[\catrequirements
]\:{\catconfiguration}\:\caterror \colsep{\triangleq}\caterror
\nonumber\\
\catsemantics[\catrequirements
]\:{\catconfiguration}\:\quadruple{j}{f}{\catenvironment}{\communicationrelationidentifiers}\colsep{\triangleq}\iif\ j=\catforbidden\ \ithen \quadruple{j}{f}{\catenvironment}{\communicationrelationidentifiers}\nonumber\\
\ielse
\mmatch\ \catrequirements
\ \mwith\nonumber\\
\mcase\catinstruction\mthen{\{\catsemantics[\catinstruction]\:\catconfiguration\:\catinstructionresultof{\catallowed}{f}{\catenvironment}{\communicationrelationidentifiers}{}\}}\nonumber\\[1ex]
\mcase\catinstruction \ \catrequirements
'\mthen\nonumber\\
\quad\llet\ r = \catsemantics[\catinstruction]\:\catconfiguration\:\catinstructionresultof{\catallowed}{f}{\catenvironment}{\communicationrelationidentifiers}\ \lin\ \nonumber\\
\quad\quad\iif\ (r=\caterror)\ \ithen\ \caterror\ \nonumber\\
\quad\quad\ielse\ \llet\ \catinstructionresultof{j}{f'}{\catenvironment'}{\communicationrelationidentifiers'}=r\ \lin\ \nonumber\\
\quad\quad\quad\iif\ (j=\catallowed)\ \ithen\nonumber\\
\quad\quad\quad\quad\catsemantics[\catrequirements']\:\catconfiguration\:\quadruple{\catallowed}{f\cup f'}{\catenvironment'}{\communicationrelationidentifiers'}\nonumber\\
\quad\quad\quad\ielse\ \:{\{r\}}
\nonumber\\[1ex]
\mcase\defcatwithtt\  \nonterminal{id}\  \catfromtt\  \catexpr{}\ \catrequirements
'\mthen\nonumber\\
\quad\llet\ \cattype{type}S\ =\ \catsemantics[\catexpr{}]\:\catconfiguration\:\catenvironment\ \lin\nonumber\\
\quad\quad\iif\ (\cattypenocolon{type}{}\neq\cattypenocolon{set}{})\ \ithen\ \caterror\nonumber\\
\quad\quad\ielse\bigcup_{e\in S}\catsemantics[\catrequirements
']\:\catconfiguration\:\quadruple{\catallowed}{f}{\catassign{\catenvironment}{\nonterminal{id}}{e}}{\communicationrelationidentifiers\cup\{\nonterminal{id}\}}\nonumber
\end{eqntabular}
\end{pcframed}
\vspace*{-2mm}
\caption{Semantics of requirements~\label{fig:spec-requirements}}\label{fig:bind-with}
\end{figure}


%
%
%
%

\newcommand{\domain}{\mathsffont{dom}}
\newcommand{\codomain}{\mathsffont{rng}}
\newcommand{\field}{\mathsffont{fld}}

\clearpage
\section{\cat{} library functions}\label{sec:cat-primitives-and-library-functions}
For reference, we give 
the code of three library functions that
operate over relations and sets (\catfold, \catmap and \catcross).


\subsection{Definition of \protect\catfold}\label{sec:def:fold}
\label{def:fold}Given a function~$f$, a set~$S=\{e_1, e_2, \dots,e_n\}$
and an element~$y$, the call \defcatfold\texttt{\ $f$\ ($S$, $y$)}
returns the value $f(e_{i_1} , f(e_{i_2}, \dots f(e_{i_n},y)))$,
where $i_1, i_2, \dots, i_n$ is a permutation of $1, 2, \dots, n$:
\begin{verbatim}
let fold f =
  let rec fold_rec (es,y) = match es with
  || {} -> y
  || e ++ es -> fold_rec (es, f(e,y))
  end in
  fold_rec
\end{verbatim}

\subsection{Definition of \protect\catmap}\label{sec:def:map}

\label{def:map}Given a function $f$ and a set $S = \{e_1, \dots, e_n\}$,
the call  \defcatmap\texttt{~$f$~$S$} returns
the set $\{f(e_1), \dots, f(e_n)\}$. This function can
be implemented directly or more concisely
by calling the \catfold function:
\begin{verbatim}
let map f = fun es -> fold (fun (e,y) -> f e ++ y) (es,{})
\end{verbatim}

\subsection{Definition of \protect\catcross}\label{sec:def:cross}

\label{def:cross}The function \catcross\ takes a set of sets $S
= \{S_1, S_2, \ldots, S_n\}$ as argument and returns all possible unions built
by picking elements from each of the $S_i$: $$ \left\{\, e_1 \cup e_2 \cup
\cdots \cup e_n \mid e_1 \in S_1, e_2 \in S_2, \ldots, e_n \in S_n \,\right\}
$$
Note that if $S$ is empty, then \catcross should
return one relation exactly: the empty relation $\emptyset$, \ie, the neutral
element of the union operator.
This choice for \catcross\texttt{($\emptyset$) = $\emptyset$} is natural
when we define \catcross\ inductively:
$$
\defcatcross(S_1 \mathop{\texttt{++}} S) =
\bigcup_{e_1 \in S_1, t \in \catcross(S)} \left\{ e_1 \cup t \right\}
$$
In this specification, we simply build
\catcross\texttt{($S_1 \mathop{\texttt{++}} S$)} by building the set
of all unions of one relation~$e_1$ picked in~$S_1$
and of one relation~$t$ picked in $\catcross(S)$.
From this inductive specification for \catcross,
one writes the following concise code:
\begin{verbatim}
let rec cross S = match S with
  || {} -> { 0 }
  || S1 ++ S ->
      let yss = cross S in
      fold
        (fun (e1,r) -> map (fun t -> e1 | t) yss | r)
        (S1,{})           
   end      
\end{verbatim}


\begin{thebibliography}{11}
\providecommand{\natexlab}[1]{#1}
\providecommand{\url}[1]{\texttt{#1}}
\expandafter\ifx\csname urlstyle\endcsname\relax
  \providecommand{\doi}[1]{doi: #1}\else
  \providecommand{\doi}{doi: \begingroup \urlstyle{rm}\Url}\fi

\bibitem[Aczel(1988)]{Aczel-88-non-wellfounded-sets}
P.~Aczel.
\newblock \emph{Non-well-founded sets}, volume~14 of \emph{CSLI Lecture Notes}.
\newblock Stanford University, Center for the Study of Language and
  Information, 1988.

\bibitem[Alglave and Cousot(2016)]{LISA-arxiv16}
J.~Alglave and P.~Cousot.
\newblock Syntax and analytic semantics of {LISA}.
\newblock \emph{CoRR}, abs/1608.06583, 2016.
\newblock URL \url{http://arxiv.org/abs/1608.06583}.

\bibitem[Alglave and Maranget(2015)]{herd7}
J.~Alglave and L.~Maranget.
\newblock \textsf{\fontsize{7}{10}\selectfont herd7}.
\newblock
  \href{http://virginia.cs.ucl.ac.uk/herd}{\url{virginia.cs.ucl.ac.uk/herd}},
  31 Aug. 2015.

\bibitem[Alglave et~al.(2015{\natexlab{a}})Alglave, Batty, Donaldson,
  Gopalakrishnan, Ketema, Poetzl, Sorensen, and Wickerson]{abd15}
J.~Alglave, M.~Batty, A.~F. Donaldson, G.~Gopalakrishnan, J.~Ketema, D.~Poetzl,
  T.~Sorensen, and J.~Wickerson.
\newblock {GPU} concurrency: Weak behaviours and programming assumptions.
\newblock In \emph{ASPLOS}, 2015{\natexlab{a}}.

\bibitem[Alglave et~al.(2015{\natexlab{b}})Alglave, Cousot, and
  Maranget]{AlglaveCousotMaranget-cat-HSA-2015}
J.~Alglave, P.~Cousot, and L.~Maranget.
\newblock Syntax and semantics of the \cat language.
\newblock \emph{HSA Foundation}, Version 1.1:\penalty0 38 p., 16 Oct
  2015{\natexlab{b}}.
\newblock URL \url{http://www.hsafoundation.com/?ddownload=5382}.

\bibitem[Batty et~al.(2016)Batty, Wickerson, and
  Donaldson]{DBLP:journals/corr/WickersonBD15}
M.~Batty, J.~Wickerson, and A.~F. Donaldson.
\newblock Overhauling {SC} atomics in {C11} and {O}pen{CL}.
\newblock In \emph{POPL}, 2016.

\bibitem[{HSA~Foundation}(2015)]{HSA-Foundation-PSAS-2015}
{HSA~Foundation}.
\newblock Hsa platform system architecture specification 1.0.
\newblock
  \href{http://www.hsafoundation.com/?ddownload=4944}{\texttt{HSA-SysArch-1.01.pdf}},
  \href{http://www.hsafoundation.com/?ddownload=5381}{\texttt{cat\_ModelExpressions-1.1.pdf}},
  15 Jan. 2015.

\bibitem[Lamport(1979)]{lam79}
L.~Lamport.
\newblock How to make a multiprocessor computer that correctly executes
  multiprocess programs.
\newblock \emph{IEEE Trans. Computers}, 28\penalty0 (9):\penalty0 690--691,
  1979.

\bibitem[Leroy et~al.(2014)Leroy, Doligez, Frisch, Garrigue, R\'emy, and
  Vouillon]{OCaml14}
X.~Leroy, D.~Doligez, A.~Frisch, J.~Garrigue, D.~R\'emy, and J.~Vouillon.
\newblock {The OCaml system, release 4.02, Documentation and user's manual }.
\newblock
  \href{http://caml.inria.fr/pub/docs/manual-ocaml/}{\url{caml.inria.fr}}, 24
  Sept. 2014.

\bibitem[Milner and Tofte(1991)]{DBLP:journals/tcs/MilnerT91}
R.~Milner and M.~Tofte.
\newblock Co-induction in relational semantics.
\newblock \emph{Theor. Comput. Sci.}, 87\penalty0 (1):\penalty0 209--220, 1991.

\bibitem[Tarski(1955)]{Tarski55-fp}
A.~Tarski.
\newblock A lattice theoretical fixpoint theorem and its applications.
\newblock \emph{Pacific J.\ of Math.}, 5:\penalty0 285--310, 1955.

\end{thebibliography}
\end{document}